\documentclass[aps,superscriptaddress,preprint,showpacs,showkeys,amsmath,amssymb,nofootinbib]{revtex4}

\usepackage{graphicx}
\usepackage{epstopdf} 

\begin{document}



\title[Combinatorial Information Theory I]{Combinatorial Information Theory: \\ I. Philosophical Basis of Cross-Entropy and Entropy}
\author{Robert K. Niven}
\email{r.niven@adfa.edu.au}
\affiliation{School of Aerospace, Civil and Mechanical Engineering, The University of New South Wales at ADFA, Northcott Drive, Canberra, ACT, 2600, Australia.}
\affiliation{Niels Bohr Institute, Copenhagen University, Denmark.}

\date{April 2007}

\begin{abstract}

This study critically analyses the information-theoretic, axiomatic and combinatorial philosophical bases of the entropy and cross-entropy concepts. The combinatorial basis is shown to be the most fundamental (most primitive) of these three bases, since it gives (i) a derivation for the Kullback-Leibler cross-entropy and Shannon entropy functions, as simplified forms of the multinomial distribution subject to the Stirling approximation; (ii) an explanation for the need to maximize entropy (or minimize cross-entropy) to find the most probable realization of a system; and (iii) new, generalized definitions of entropy and cross-entropy - supersets of the Boltzmann principle - applicable to non-multinomial systems. The combinatorial basis is therefore of much broader scope, with far greater power of application, than the information-theoretic and axiomatic bases.  The generalized definitions underpin a new discipline of ``{\it combinatorial information theory}'', for the analysis of probabilistic systems of any type.

Jaynes' generic formulation of statistical mechanics for multinomial systems is re-examined in light of the combinatorial approach, including the analysis of probability distributions, ensemble theory, Jaynes relations, fluctuation theory and the entropy concentration theorem. Several new concepts are outlined, including a generalized Clausius inequality, a generalized free energy (``free information") function, and a generalized Gibbs-Duhem relation and phase rule. For non-multinomial systems, the generalized approach provides a different framework for the reinterpretation of the many alternative entropy measures (e.g.\  Bose-Einstein, Fermi-Dirac, R\'{e}nyi, Tsallis, Sharma-Mittal, Beck-Cohen, Kaniadakis) in terms of their combinatorial structure. A connection between the combinatorial and Bayesian approaches is also explored.

\end{abstract}





\maketitle

%
%
%

\section{\label{sect1}Introduction} 
%
The concept of {\it entropy} - a measure of the lack of order of a system - is one of the most profound human discoveries, with implications for virtually all disciplines of human study. The {\it thermodynamic entropy} was first defined by Clausius \cite{Clausius_1865} in terms of the exact differential $dS$, given by the quantity of heat transferred reversibly to a system, $dQ$, scaled by the absolute temperature $T$ of the system:
\begin{equation}
dS = \frac{dQ}{T}
\label{eq1}
\end{equation}
Consideration of irreversible (non-equilibrium) processes - as expressed by the second law of thermodynamics - gives the Clausius \cite{Clausius_1876} inequality:
\begin{equation}
dS \geq \frac{dQ}{T}
\label{eq2}
\end{equation}
The {\it combinatorial basis} of entropy was revealed by Boltzmann \cite{Boltzmann_1877} and Planck \cite{Planck_1901, Planck_1913} in the famous equation: 
\begin{equation}
S_N  = NS = k\ln \mathbb{W}
\label{eq3}
\end{equation}
where $S_N$ is the total thermodynamic entropy of the system, $N$ is the number of entities (discrete particles or agents) present, $S$ is the thermodynamic entropy per entity, $\mathbb{W}$ is number of ways in which a specified realization\footnote{Here the {\it state} refers to each different category (e.g.\  boxes, energy levels, elements or results) accessible to a system; a {\it configuration} is a distinguishable permutution or pattern of entities amongst the states of a system ({\it complexion}, {\it microstate}, {\it sequence}); and a {\it realization} is an externally identifiable set of such configurations, grouped in a specified manner ({\it macrostate}, {\it outcome}, {\it type}).} of a system can occur, known as its statistical weight, and $k$ is the Boltzmann constant ($1.38\times10^{23}~{\rm J~K^{-1}~entity^{-1}}$).  Whilst the combinatorial definition is well-known in physics, most thermodynamicists, information theorists and mathematicians use the {\it information entropy} of Shannon \cite{Shannon_1948} (or a simple multiple thereof, as given by Boltzmann \cite{Boltzmann_1877}):
\begin{equation}
H({\mathbf{p}}) =  \frac{S_N}{kN} = - \sum\limits_{i = 1}^s {p_i \ln p_i } 
\label{eq5}
\end{equation}
where $p_i$ is the probability of occurrence of the $i$th distinguishable outcome or state, from $s$ such states, and ${\mathbf p}\!=\!\{p_i\}$ is the probability distribution (probability mass function).  In the {\it maximum entropy principle} (``MaxEnt'') developed by Jaynes \cite{Jaynes_1957}, (\ref{eq5}) is maximized subject to the constraints on a system, giving a probability distribution ${\mathbf p}^*$ which is considered to be the most uncertain, the least informative or the least committed to information not given, and is therefore used to represent the system \cite{Jaynes_1957, Tribus_1961a, Tribus_1961b, Jaynes_1963, Jaynes_1968, Jaynes_1978, Kapur_K_1987, Kapur_K_1992, Jaynes_2003}.  It has been shown that the central tenets of thermodynamics can be derived directly (and more naturally) using the MaxEnt principle without recourse to any other laws \cite{Jaynes_1957, Tribus_1961a, Tribus_1961b}.  The Shannon entropy is itself a subset of the Kullback-Leibler {\it directed divergence}, {\it cross-entropy} or {\it relative entropy}\
\!\footnote{The relative entropy is usually defined as the negative of (\ref{eq6}); a few authors define the cross-entropy differently.} 
function \cite{Kullback_L_1951, Kullback_1959, Jaynes_1963, Snickars_W_1977, Kapur_K_1992}, in discrete form:
\begin{equation}
D({\mathbf{p}}|{\mathbf{q}}) = \sum\limits_{i = 1}^s {p_i \ln \frac{{p_i }}{{q_i }}} 
\label{eq6}
\end{equation}
where $q_i$ and $p_i$ are respectively the prior and posterior probability of occurrence of the $i$th result, ${\mathbf p}=\{p_i\}$, ${\mathbf q}=\{q_i\}$ and the solidus $|$ is the Bayesian ``subject to".  In the {\it minimum cross-entropy principle} (``MinXEnt'') - a superset of MaxEnt - (\ref{eq6}) is minimized subject to the constraints on a system, to give the distribution ${\mathbf p}^*$ which contains the least information, yet is closest to ${\mathbf q}$ \cite{Kapur_K_1992}.  Collectively, definitions (\ref{eq1})-(\ref{eq6}) underpin present-day statistical physics, thermodynamics, information theory and encoding, optimization and data analysis, whilst the MaxEnt and MinXEnt methods have found widespread application to a vast assortment of fields, including information technology, communications, mathematics, science, engineering, economics, decision theory, geography, linguistics and the social sciences \citep[e.g.][]{Tribus_1961b, Tribus_1969, Kapur_K_1987, Kapur_1989b, Kapur_K_1992}.  

Attention must, however, be directed towards two points.  Firstly, although the MaxEnt and MinXEnt methods are widely considered to fall within the scope of Bayesian inferential reasoning \citep[e.g.][]{Tribus_1988, Jaynes_1988b}, no such Bayesian derivation appears to be evident in the literature, and the two methods remain largely underpinned by axiomatic arguments \cite{Shore_J_1980, Levine_1980}.  Since these arguments can be varied and are subject to debate, the circumstances in which the MaxEnt and MinXEnt methods remain valid, or require modification, are not clear.  Secondly, there is now widespread controversy in many of the above-mentioned fields - especially in statistical physics - due to the promulgation of a variety of alternative entropy functions, inconsistent with the above definitions, for specific applications. These include the Fisher information measure \cite{Fisher_1922, Fisher_1925}; the Bose-Einstein and Fermi-Dirac entropies of quantum mechanics \cite{Bose_1924, Einstein_1924, Einstein_1925, Fermi_1926, Dirac_1926, Tolman_1938, Davidson_1962}; the R\'{e}nyi \cite{Renyi_1961}, Tsallis \cite{Tsallis_1988, Tsallis_2001}, Sharma-Mittal \cite{Sharma_M_1975, Sharma_M_1977} and other entropies of non-extensive (correlated) statistics; Beck-Cohen superstatistics \cite{Beck_C_2003}; the Kaniadakis entropy of relativity theory \cite{Kaniadakis_2001, Kaniadakis_2002}; the ``exact'' entropies of the author \cite{Niven_2005, Niven_2006}, and many others \citep[e.g.][]{Aczel_1975, Burbea_1983, Papaioannou_1985, Kapur_1983, Kapur_1984, Kapur_1986, Kapur_1989b, Behara_1990, Arndt_2001}. In recent years, there has been a tremendous surge of interest in such alternative measures; e.g.\  the Tsallis (non-extensive) literature alone contains over 1000 refereed journal articles since 1988, with 141 in 2005. Despite this high level of activity, the fundamental meaning of such alternative entropy functions, and how they fit into the combinatorial scheme of Boltzmann, is still not well understood. 

This discussion highlights the fact that the entropy concept has many different philosophical bases. In addition to (i) the combinatorial basis of entropy, other bases include:
\begin{enumerate}
\item[(ii)] The {\it information-theoretic basis} \cite{Szilard_1929, Shannon_1948, Wiener_1948, Brillouin_1951a, Brillouin_1953, Tribus_M_1971,Yaglom_Y_1983, Cover_T_1991}, in which entropy is defined in terms of the number of {\it bits} of information needed to describe a particular system, and/or in terms of coding theory;
\item[(iii)] The {\it axiomatic basis} \cite{Shannon_1948}, in which the desired properties of an entropy measure - its {\it axioms} - are listed and used for its derivation;
\item[(iv)] The {\it inverse modelling approach} of Kapur and Kesavan 
\cite{Kapur_K_1987, Kesavan_K_1989, Kapur_K_1992, Kapur_etal_1995, Yuan_Kesavan_1998, Srikanth_2000}, in which one works backwards from an observed probability distribution ${\mathbf p}^*$, a priori distribution ${\mathbf q}$ (if available) and any constraints, to derive the measure of cross-entropy or entropy applicable to a system;
\item[(v)] The {\it game-theoretic basis} \cite{vonNeumann_M_1944, Nash_1950, Topsoe_1993, Harremoes_T_2001, Topsoe_2002, Topsoe_2004, Grunwald_2004}, in which an entropy function is derived by analysis of a game between two or more players; and
\item[(vi)] The {\it information-geometric} or {\it statistical manifold basis} \cite{Bhattacharyya_1943, Rao_1945, Amari_1985, Burbea_1986}, in which an information measure is analysed using a geometric representation. 
\end{enumerate}
Bases (ii) and (iii) are popular in information theory, (v) in economics, business and military strategy, (vi) in statistics, probability theory and mathematics, whilst (iv) is less well known. Whilst each basis has a following in its own discipline, the relationships between the different bases are still largely unexplored. Furthermore, whether any basis can claim supremacy over the other bases, or whether they are of equal philosophical standing, is a question which has not been adequately addressed.
  
The aim of this article 
- which follows two previous studies \cite{Niven_2005, Niven_2006} - is to critically examine the philosophical bases of the entropy and cross-entropy concepts, with particular attention to the information-theoretic, axiomatic and combinatorial interpretations.  Using the combinatorial basis, it is shown (following a well-trodden road) that both the cross-entropy and entropy functions are simplified forms of the logarithm of the multinomial distribution; they are therefore only shorthand functions to determine the ``most probable'' (MaxEnt or MinXEnt) realization of a system which follows the multinomial distribution, without the necessity of invoking this distribution itself.  The Kullback-Leibler cross-entropy and Shannon entropy functions are therefore secondary concepts, based firmly on simple combinatorial principles.  This perspective lies in stark contrast to the axiomatic and information-theoretic bases, both of which take the cross-entropy or (especially) the entropy function as the fundamental concept and starting point for analysis. Since it rests upon a more definitive philosophical foundation, the combinatorial basis is the most fundamental (most primitive) of these three bases.  It is also of much broader scope, leading naturally to new, {\it generalized combinatorial definitions} of cross-entropy and entropy - each a superset of the Boltzmann principle (\ref{eq1}) - for the analysis of {\it any} probabilistic system, irrespective of whether it is governed by the multinomial distribution.  (Such definitions permit the {\it reinterpretation} of the many alternative cross-entropy and entropy measures - e.g.\  Bose-Einstein, Fermi-Dirac, R\'{e}nyi, Tsallis, Sharma-Mittal, Beck-Cohen, Kaniadakis, etc - in light of their combinatorial structure.)  The revised definitions underpin the development of a new, broad discipline of {\it combinatorial information theory}, spanning the entirety of present-day statistical physics, information theory and probability theory, for the analysis of probabilistic systems of every type.

After early drafts of this work were completed \cite{Niven_2006_benhur}, the author's attention was alerted to several works of Grendar and Grendar \cite{Grendar_G_2000, Grendar_G_2001, Grendar_G_2004a, Grendar_G_2004b, Grendar_2004c}, who adopt a similar philosophical argument, albeit with somewhat different aims and a different scope.  In fact, the central premise of this study has been known since the time of Boltzmann \cite{Boltzmann_1877}, played a critical role in the discovery of Bose-Einstein and Fermi-Dirac (quantum) statistics \cite{Bose_1924, Einstein_1924, Einstein_1925, Fermi_1926, Dirac_1926}, and to some extent provides a motivation for present-day large deviations theory \citep[e.g.][]{Ellis_1985, Dembo_Z_1993}, but for some reason has not been developed to its logical conclusion, {\it viz.}\ into generalized combinatorial definitions of entropy and cross-entropy.  The study therefore encompasses and expands upon the combinatorial arguments used in classical and quantum statistical mechanics \citep[e.g.][]{Boltzmann_1877, Planck_1901, Planck_1913, Tolman_1938, Brillouin_1951b, Schrodinger_1952, Hill_1956, Davidson_1962, Eyring_1964, Desloge_1966, Abe_1975, McQuarrie_1976, Atkins_1982}.  Such arguments tend to be examined only in passing by most information theorists, although there are some notable exceptions \citep[e.g.][]{Jaynes_1963, Jaynes_1968, Jaynes_2003, Snickars_W_1977, Jefferson_S_1986, Kapur_1989a}.  

This work is organised as follows. In \S\ref{sect2_1}-\ref{sect2_3}, the main elements of the information-theoretic, axiomatic and combinatorial bases of entropy and cross-entropy are critically examined, leading to combinatorial derivations of the Shannon and Kullback-Leibler measures, which reveal their purpose to determine the ``most probable'' (modal) probability distribution of a multinomial system.  Several technical aspects are then scrutinized in detail: zero reference states for entropy or information; ensemble theory and multicomponent systems; and the ``generic'' formulation of statistical mechanics developed by Jaynes \citep[e.g.][]{Jaynes_1957, Jaynes_1963, Jaynes_1968, Jaynes_1978, Jaynes_2003, Tribus_1961a, Tribus_1961b, Kapur_K_1987, Kapur_K_1992}.  The latter is reinterpreted and extended for a multinomial system in light of the combinatorial approach, with the derivation of new concepts including a generalized Clausius inequality, a generalized free energy (``free information") function, a generalized Gibbs-Duhem relation and phase rule, and a reappraisal of fluctuation theory and Jaynes' entropy concentration theorem. In \S\ref{sect3}, the significance of the multinomial distribution is reviewed, leading to the proposition of generalized definitions of entropy and cross-entropy for non-multinomial systems.  A connection to Bayesian statistical inference, and the other bases of entropy, are also discussed.


In the following, an {\it entity} is taken to be any discrete particle, object or agent within a system, which acts separately but not necessarily independently of the other entities present (note this definition encompasses human beings).  The entity therefore constitutes the unit of analysis of the system, although of course some entities can be further examined in terms of their constituent sub-entities, if desired. 

\section{\label{sect2}Theoretical Roots of the Information Entropy Concept} 
%
What is entropy?  This question has certainly occupied (or been dismissed from) the minds of millions of college and university students for one and a half centuries - predominantly in physics, chemistry, engineering and informatics - and undoubtedly tens of thousands more of their professional elders in all disciplines.  To endeavour to answer this question, in this section the first three theoretical or philosophical roots of the entropy and cross-entropy concepts listed in \S\ref{sect1} are examined.  The first two, information-theoretic and axiomatic, are so closely intertwined in the literature that it is not possible to distinguish them clearly.  The third origin, based on combinatorial analysis, is somewhat distinct, and occupies much of this work.  Discussion of the remaining three bases of entropy (inverse modelling, game-theoretic and information-geometric) is postponed until later in the text (\S \ref{sect3_3}).  A rival approach to the analysis of probabilistic systems, which invokes the continuous Fisher information \cite{Fisher_1922, Fisher_1925, Frieden_2004}, is examined in detail elsewhere \cite{Niven_2006_Fisher}. 

\subsection{\label{sect2_1}The Information-Theoretic (Bits) Approach} 
The first theoretical basis of the Shannon entropy - although not the first in historical development - concerns the number of bits of information required to specify a particular system or outcome \cite{Szilard_1929, Shannon_1948, Wiener_1948, Brillouin_1951a, Brillouin_1953, Tribus_M_1971,Yaglom_Y_1983, Cover_T_1991}.  Consider the {\it binary entropy} or $B$-entropy:
\begin{equation}
B =  - \sum\limits_{i = 1}^s {p_i \log _2 p_i } 
\label{eq7}
\end{equation}
related to the Shannon entropy (defined using the natural logarithm, (\ref{eq5})) by $H = B \ln2$.  Now consider a random variable which may take one of two states, of equal probability $p_i=\tfrac{1}{2}, i=1,2$.  Initially, the state of the variable is not known.  After a {\it binary decision} (a process of selection or measurement) it is found to be in one of these states (say $p_1=1$) and not the other ($p_2=0$).  The initial and final binary entropies are therefore:
\begin{equation}
B_{init} = -2(\tfrac{1}{2} \log_2 \tfrac{1}{2})=1, \qquad B_{final} = -(1\log_2 1 + 0\log_2 0) = 0
\label{eq8}
\end{equation}
(Here and subsequently, we take $0\log0=\log0^0=\log1=0$ for all logarithmic bases).  The change in entropy is then:
\begin{equation}
\Delta B = B_{final} - B_{init} =  -1
\label{eq9}
\end{equation}
If we {\it define} the change in information as the negative of the change in entropy (i.e., entropy lost = information gained) \cite{Schrodinger_1952, Wiener_1948, Brillouin_1949, Brillouin_1950, Brillouin_1951a, Brillouin_1953}, the gain in information - reflecting our improved state of knowledge - is:
\begin{equation}
\Delta I =  -\Delta B = 1
\label{eq10}
\end{equation}
Thus for a simple binary decision, the information gained (entropy lost) corresponds to one {\it bit} of information.  The decrease in entropy therefore provides a quantitative measure of the information gained by observation of a system.

If we adopt a scaled binary entropy $S_B = -k\sum\nolimits_{i = 1}^s {p_i \log_2 p_i}$, the information gained by a binary decision is $k$, measured in the units of $k$.  For a scaled entropy based on the natural logarithm, $S= -k\sum\nolimits_{i = 1}^s {p_i \ln p_i }$, the gain in information is $k \ln2$ \cite{Szilard_1929, Shannon_1948}.  For thermodynamic systems for which $k$ is the Boltzmann constant, 1 bit of information corresponds to an energy transfer of $9.57 \times 10^{-24} {\rm~J~K^{-1}~entity^{-1}}$.  To access information carried by photons, and distinguish them from the background (thermal) radiation, it is necessary to account for the effect of temperature \cite{Brillouin_1951a, Brillouin_1953}; in this case, 1 bit of information corresponds to $kT \ln2$ energy units per entity.

A second variant of the information-theoretic definition - which overlaps with the axiomatic approach (\S \ref{sect2_2}) - is to consider a random variable which may take $s$ equally probable states.  We define a measure of {\it uncertainty} as \cite{Hartley_1928, Tribus_1961b}:
\begin{equation}
U = \ln s
\label{eq11}
\end{equation}
As the states are equally probable, $s=1/p_i, \forall i$, hence $U=-\ln p_i$.  The mathematical expectation of the uncertainty is $\left\langle U \right\rangle = -\sum\nolimits_{i = 1}^s {p_i \ln p_i } = H$, i.e.\ the Shannon entropy.  As the states are equally probable, this reduces to $\left\langle U \right\rangle  = U$.
  
For states which are not equally probable, we may thus adopt the Shannon entropy as a measure of the expectation of the uncertainty \cite{Shannon_1948}.  We can further define the {\it surprisal} or {\it self-information} associated with each result \cite{Tribus_1961a, Tribus_1961b, Burbea_1983}:
\begin{equation}
\sigma_i  =  - \ln p_i
\label{eq12}
\end{equation}
The entropy is therefore the expectation of the surprisal.  

The surprisal has also been defined relative to the prior probability of that result, $\delta_i = \ln (p_i /q_i )$, i.e.\ as the amount of information gained by a decision or message \cite{Kullback_L_1951, Tribus_1961b, Burbea_1983}.  This is better referred to as the {\it cross-surprisal}.  The expectation of the cross-surprisal gives the cross-entropy (\ref{eq6}).  The cross-entropy is therefore a measure of the expected information relative to what is known.  Another useful term is the function $H_i=-p_i \ln p_i$, here termed the {\it weighted surprisal} or {\it partial entropy}, which when summed over all states gives the Shannon entropy \citep[c.f.][]{Young_1971, Yaglom_Y_1983, Cover_T_1991, Niven_2004}.  The analogous function $D_i  = p_i \ln (p_i /q_i)$ can be termed the {\it weighted cross-surprisal} or {\it partial cross-entropy}.  

The third and strongest variant of the information-theoretic approach relates to information coding \citep[e.g.][]{Cover_T_1991}, in which an alphabet $\mathbb{A}=\{a_i\}$ with known or inferred probabilities $\{p_i\}$ is mapped to a binary code\!\footnote{In general, $\mathbb{A}$ can be mapped to a code alphabet $\mathbb{K}=\{k_i\}$ of any size \cite{Cover_T_1991}.}, with corresponding codeword lengths $ \{\kappa_i\}, \kappa_i \in \mathbb{N}, \forall i$. To minimize the mean codeword length, we consider the binary entropy:
\begin{equation}
B_0 =  \min_{\kappa \in \text{all codes}}  \sum\limits_{i = 1}^s {p_i \kappa_i}. 
\label{eqB0}
\end{equation}
To obtain an {\it instantaneous} or readily decipherable code, it is common practice to seek a {\it prefix-free} code, in which no codeword is a prefix of any other; the codeword lengths are then subject to the Kraft inequality \cite{Cover_T_1991}:
\begin{equation}
\sum\limits_{i = 1}^s {2^{-\kappa_i} \leq 1 }. 
\label{eqKraft}
\end{equation}
Minimization of (\ref{eqB0}) with respect to $\kappa_i$ subject to (\ref{eqKraft}) by the Lagrangian method (see \S \ref{sect2_3_2}), with normalization ($\sum_{i=1}^s {p_i} =1$), yields a discontinuous binary entropy:
\begin{equation}
B_0 =  \sum\limits_{i = 1}^s { p_i  \lceil{- \log _2 p_i}\rceil } 
\label{eqB0quant}
\end{equation}
where $\lceil x \rceil$ is the ceiling of $x$ (the smallest integer greater than or equal to $x$), which arises since $\kappa_i$ must be an integer. By repeated $m$-fold sampling of (\ref{eqB0quant}), the two entropies converge:
\begin{equation}
B =  \lim_{m \to \infty} \frac {B_0}{m} 
\label{eqBconv}
\end{equation}
The entropy $B$ therefore indicates the minimum mean (possibly fractional) number of bits per symbol, whilst 
 $B_0$ is the equivalent quantity based on integer codeword lengths.

The above three information-theoretic roots of the Shannon entropy are of tremendous utility, primarily to information theory and coding applications. However, the first two variants suffer from the deficiency that they {\it assume} that measures of information (or entropy) should be of logarithmic form, an assumption in part derived from the axiomatic approach (\S \ref{sect2_2}).  Certainly, other functions could yield one bit of information for a binary decision (\ref{eq10}). The third variant assumes that the mean code length is the appropriate quantity to be minimized; this is reasonable for coding applications, but does not necessarily apply to other situations. Furthermore, the Kraft inequality - which gives rise to the logarithm in the binary entropy - is not universal in application (e.g.\ to fixed-length codes, codes incorporating redundancy, etc), and warrants further examination. In consequence, the information-theoretic definitions of entropy and cross-entropy have a narrow philosophical basis, which does not necessarily apply outside their domain of application.


\subsection{\label{sect2_2}The Axiomatic Approach} 
The second theoretical basis of the entropy concept, developed by Shannon \cite{Shannon_1948}, proceeds by listing the desired properties of a measure of uncertainty - its {\it axioms} or {\it desiderata} - and finding the mathematical function which satisfies these axioms.  Shannon \cite{Shannon_1948} considered three axioms: continuity, monotonicity and recursivity (the branching principle), from which the Shannon entropy (\ref{eq5}) is uniquely obtained.  To Shannon's original list, many additional axioms have been added: e.g.\  uniqueness, permutational symmetry (invariance), non-negativity, non-impossibility, inclusivity, decisivity, concavity, maximum entropy at uniformity (normality), additivity, strong additivity, subadditivity, system independence and subset independence  \citep[e.g.][]{Shannon_1948, Tribus_1961b, Aczel_1975, Shore_J_1980, Kapur_1983, Skilling_1988, Kapur_1989b, Behara_1990, Kapur_K_1992}.  The Shannon entropy is the only function which satisfies these axioms.  Indeed, it may be deduced from several small subsets of these axioms, implying that they are not independent \citep[e.g.][]{Kapur_1983, Tribus_1988, Kapur_K_1992}.  

It must be noted that the definition of thermodynamic entropy (\ref{eq3}) by Planck \cite[\S118]{Planck_1913} is derived by an axiomatic argument, assuming multiplicity of the weights and additivity of the entropy function.  Similarly, in the ``plausible reasoning" treatises of Cox \cite[p37]{Cox_1961} and Jaynes \cite[\S2.1]{Jaynes_2003}, the Shannon entropy (\ref{eq5}) is obtained axiomatically, assuming entropy is additive and multiply differentiable.

The cross-entropy or directed divergence function $D$ can also be obtained using the axiomatic approach \cite{Kullback_L_1951, Kullback_1959, Shore_J_1980, Kapur_K_1992}.  Its governing axioms are broadly similar to those for the Shannon entropy, except that it is convex, and the equilibrium distribution ${\mathbf p^*} = {\mathbf q}$ in the absence of other constraints \cite{Kapur_K_1992}. Both the MaxEnt and MinXEnt principles themselves have also been justified axiomatically \citep[e.g.][]{Shore_J_1980, Levine_1980}.

Whilst mathematically sound and of tremendous utility, the axiomatic approach is intellectually unsatisfying in that it presents an austere, sterile basis for the entropy and cross-entropy functions, based only on abstract notions of desirable properties.  The answer to the question - what is entropy? - is still not clear.  Further, as Kapur \cite[p209]{Kapur_1983} notes: ``{\it mathematicians tried to modify these axioms to get more general measures} [of uncertainty] {\it including Shannon's measure as a special or limiting case}". Other entropy functions, which do not reduce to the Shannon entropy, have also been derived using different sets of axioms \citep[e.g.][]{Renyi_1961, Aczel_1975, Kapur_1983, Kapur_1989b, Tsallis_1988, Tsallis_2001, Behara_1990, Arndt_2001}.  Other measures of divergence have also been proposed \citep[e.g.][]{Kapur_1984, Burbea_1983, Papaioannou_1985, Arndt_2001}.  How can we be certain that the axioms used to derive the Shannon or Kullback-Leibler measures are correct?  Indeed, the specification of particular axioms may preclude the identification of different or broader measures of entropy, which may be more appropriate for particular or more general circumstances.  To resolve these circular arguments, we now turn to consideration of the combinatorial basis of the entropy and cross-entropy functions, which as will be shown, should be recognized as their primary (most primitive) philosophical basis.

\subsection{\label{sect2_3}The Combinatorial (Statistical Mechanical) Approach} 
\subsubsection{\label{sect2_3_1}Statistics of Multinomial Systems} 
The combinatorial approach was first developed in statistical thermodynamics, to examine the distribution of molecules amongst energy levels or phase space elements \citep[e.g.][]{Boltzmann_1877, Planck_1901, Planck_1913, Tolman_1938, Brillouin_1951b, Schrodinger_1952, Hill_1956, Davidson_1962, Eyring_1964, Desloge_1966, Abe_1975, McQuarrie_1976, Atkins_1982}.  However, the combinatorial basis is only touched upon by many prominent statistical mechanics texts \citep[e.g.][]{Fowler_1936} in favour of a quantum mechanical treatment, which tends to disguise its statistical foundation. The connection between combinatorial concepts and entropy is not prominent in the information theory literature, although there are a number of notable exceptions \citep[e.g.][]{Jaynes_1963, Jaynes_1968, Jaynes_2003, Snickars_W_1977, Jefferson_S_1986, Kapur_1989a}.

\begin{figure}
\includegraphics[width=100mm]{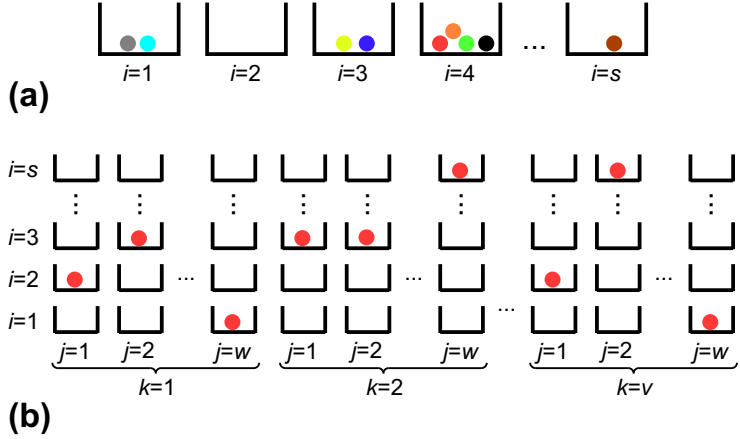}
\caption{Multinomial (a) Òballs-in-boxesÓ and (b) Òmultiple selectionÓ systems.}
\label{fig:ballsboxes}
\end{figure}

Consider the ``balls-in-boxes" system illustrated in Figure \ref{fig:ballsboxes}a, in which $N$ distinguishable balls or entities are distributed amongst $s$ distinguishable boxes or states.  This may be taken to represent $N$ molecules amongst $s$ energy levels, phase space elements or eigenfunctions\ 
\!\footnote{The boxes are here taken to be discrete, although there is no conceptual difficulty in generalizing the analysis to boxes of infinitesimal spacing.  Similarly, the number of states $s$ is considered finite, but the limit $s \to \infty$ can be considered if handled carefully \cite{Jaynes_2003}.}\!
; $N$ ensemble members amongst $s$ ensemble energy values; $N$ people amongst $s$ shops; $N$ cars amongst $s$ floors of a parking station, and so on.  We consider each realization of the system, defined to contain $n_1$ balls in box 1, $n_2$ balls in box 2, etc, or in general $n_i$ balls in box $i$.  The $N$ balls are taken to be distinguishable, but their permutations within each box are indistinguishable, i.e. we can only (or need only) distinguish the balls within any given box from those in the other boxes.  Each choice (of a ball in a box) is assumed independent of the other selections.  The probability of any particular realization of the system, $\mathbb{P}$ (equal to the probability that there are $n_i$ balls in the $i$th box, for each $i$), is given by the {\it multinomial distribution} \cite{de_Moivre_1712, Feller_1957, Ratnaparkhi_1985}:
\begin{equation}
\mathbb{P}=\mathbb{P}({\mathbf{n}}|{\mathbf{q}},N,s) = \frac{{N!}}
{{n_1 !n_2 !...n_s !}}q_1 ^{n_1 } q_2 ^{n_2 } ...q_s ^{n_s }  = N! \prod\limits_{i = 1}^s {\frac{{q_i ^{n_i } }}
{{n_i !}}} 
\label{eq13}
\end{equation}
where again $q_i$ is the prior probability of a ball falling in the $i$th box, and ${\mathbf{n}}=\{n_i\}$.  If the prior distribution ${\mathbf q}$ is equated to the uniform distribution ${\mathbf u}$ (i.e. $q_i=u=1/s, \forall i$) this reduces to:
\begin{equation}
\mathbb{P}_\mathbf{u}=\mathbb{P}({\mathbf{n}}|{\mathbf{u}},N,s) = \frac{{N!}}{{\prod\limits_{i = 1}^s {n_i !} }}s^{ - N} 
\label{eq14}
\end{equation}
Since the total number of configurations of a multinomial distribution is $s^N$ \cite{Buteo_1559}, the number of ways in which any particular realization in (\ref{eq14}) can be produced, or its statistical {\it weight}, is \cite{Bhaskara_1150, IMDMI_1636}:
\begin{equation}
\mathbb{W} = \mathbb{P}_\mathbf{u}\, s^N  = \frac{{N!}}{{\prod\limits_{i = 1}^s {n_i !} }}
\label{eq15}
\end{equation}
For constant $N$, the above equations are subject to the {\it natural constraint}:
\begin{alignat}{3}
&{\rm C0}:& \qquad \sum\limits_{i = 1}^s {n_i }  &= N&
\label{eq16}
\end{alignat}
and usually one or several {\it moment constraints} \citep[c.f.][]{Jaynes_1957}:
\begin{alignat}{3}
&{\rm C1~to~C}R:& \qquad \sum\limits_{i = 1}^s {n_i f_{ri} } &= N\left\langle {f_r} \right\rangle, &\qquad r=1,...,R
\label{eq17}
\end{alignat}
where $f_{ri}$ is the value of the function $f_r$ in the $i$th state and $\left\langle {f_r} \right\rangle$ is the mathematical expectation of $f_{ri}$.  An example of (\ref{eq17}) is an energy constraint, in which each state is of energy $f_{1i}= \varepsilon_i$ and the expectation of the energy is $\left\langle {f_1} \right\rangle = \left\langle \varepsilon \right\rangle$.

Now consider a sequence of $v$ independent and identically distributed (i.i.d.) probabilistic {\it events}, within each of which $w$ trials or selections are made between $s$ distinguishable states, as represented in Figure \ref{fig:ballsboxes}b.  Examples include tosses of a coin or coins, throws of a die or dice, spins of a roulette wheel, choices of symbols to make up a communications signal, or the sexual liaisons of leading film star.  So long as we are only interested in the statistical nature of the selections, and not their order, the probability of any realization or type (without regard to order, assuming each event is independent) also follows the multinomial distribution (\ref{eq13}) with $N=vw$.  When only one selection is made in each event (i.e.\ $w=1$), then $N=v$.  When the prior probabilities $q_i$ of each state within each selection are identical, the weight also follows (\ref{eq15}).  


\subsubsection{\label{sect2_3_2}The Most Probable Realization}  
We now use {\it first combinatorial principles} to determine the {\it most probable realization} of the multinomial systems considered.  As mentioned, the following derivation is common in statistical thermodynamics \citep[e.g.][]{Boltzmann_1877, Planck_1901, Planck_1913, Tolman_1938, Brillouin_1951b, Schrodinger_1952, Hill_1956, Davidson_1962, Eyring_1964, Desloge_1966, Abe_1975, McQuarrie_1976, Atkins_1982}, although such workers base their derivations on the weight $\mathbb{W}$.  As it is based on $\mathbb{P}$ rather than $\mathbb{W}$, the following derivation incorporates the prior distribution ${\mathbf q}$, and is therefore more comprehensive \cite{Kapur_1989a}.

Clearly, the most probable realization is that for which $\mathbb{P}$ (\ref{eq13}) is a maximum, subject to the constraints C0-C{\it R} on the system ((\ref{eq16}), (\ref{eq17})).  As the natural logarithm $\ln x$ increases monotonically with $x$, but transforms a product into a sum, it is convenient - and equivalent - to maximize $\ln \mathbb{P}$ rather than $\mathbb{P}$ itself, a convention adopted (implicitly) throughout statistical mechanics \cite{Boltzmann_1877, Atkins_1982}.  (The use of logarithms is therefore merely a matter of convenience, not a requirement.)  The most probable realization is given by:
\begin{equation}
d\left( \ln \mathbb{P}| \text{constraints} \right) = 0
\label{eq18}
\end{equation}
where $d$( ) is the total derivative or variational operator.  Now (\ref{eq18}) can be constructed using Lagrange's method of undetermined multipliers \cite{Tribus_1961b, Atkins_1982, Kapur_K_1992}, involving extremization of the Lagrangian $\mathfrak{L}$:
\begin{equation}
d\mathfrak{L} = 0
\label{eq19}
\end{equation}
From the multinomial distribution (\ref{eq13}):
\begin{equation}
\ln \mathbb{P} = \sum\limits_{i=1}^s {\left( {\frac{{n_i }}{N} \ln N! - \ln n_i ! + n_i \ln q_i } \right)} 
\label{eq20}
\end{equation}
in which (for reasons which will be explained) the leading $\ln N!$ term is brought inside the summation using the natural constraint (\ref{eq16}).  From the constraints (\ref{eq16})-(\ref{eq17}), the Lagrangian is:
\begin{equation}
\begin{split}
\mathfrak{L}  &= \sum\limits_{i=1}^s {\left( {\frac{n_i}{N}\ln N! - \ln n_i ! + n_i \ln q_i } \right)} 
 - (\lambda_0  - 1) \left( {\sum\limits_{i=1}^s {n_i}  - N} \right) \\
 & \qquad - \sum\limits_{r = 1}^R {\lambda_r \left({\sum\limits_{i = 1}^s {n_i f_{ri}}  - \left\langle {f_r } \right\rangle N} \right)} 
\end{split}
\label{eq21}
\end{equation}
where $\lambda_r$, $r=0,...,R$, are the Lagrangian multipliers, and $\lambda_0-1$ is chosen rather than $\lambda_0$ for mathematical convenience.  For constant $N$, $q_i$ and $\left\langle {f_r} \right\rangle$, and for $f_{ri}$ independent of $n_i$, we need only consider the variation of (\ref{eq21}) with respect to $n_i$, i.e. $\partial \mathfrak{L}/\partial n_i \ dn_i = 0, \forall i$, whence:
\begin{equation}
\frac{1}{N}\ln N! - \frac{\partial}{\partial n_i } \ln n_i ! + \ln q_i - (\lambda_0  - 1) 
- \sum\limits_{r = 1}^R {\lambda _r f_{ri} }  = 0, \qquad i = 1,...,s
\label{eq22}
\end{equation}
The above equations are expressed in terms of $n_i$, and can be said to be in ``$n_i$ form."

At this stage the near-universal approach taken in the literature (see previous statistical mechanics references) is to employ a truncated form of the approximation for factorials derived by Stirling \cite{Stirling_1730} and de Moivre \cite{de_Moivre_1733}:
\begin{equation}
\ln x! \approx x\ln x - x
\label{eq23}
\end{equation}
This is accurate to within 1\% of $\ln x!$ for $x>90$; a more precise form, $\ln x! \approx x \ln x - x + \frac{1}{2} \ln (2 \pi x)$, is accurate to within 1\% of $\ln x!$ for $x\!>\!4$ \cite{Feller_1957}.  (Strictly speaking, the limit must be taken using upper and lower bounds, as treated in large deviations theory \cite{Ellis_1985, Dembo_Z_1993}, but for the purpose of the present study, the Stirling approximation yields the same result). Thus $\partial \ln n_i !/\partial n_i \approx \ln n_i$ and $\ln N! \approx N\ln N - N$, and so the most probable realization, here designated with an asterisk, is obtained from (\ref{eq22}) in conjunction with C0 (\ref{eq16}) as \citep[c.f.][]{Jaynes_1957, Tribus_1961a, Tribus_1961b, Kapur_1989b, Kapur_K_1992}:
\begin{equation}
n_i ^*|q_i  \approx Nq_i \exp \left( { - \lambda _0  - \sum\limits_{r = 1}^R {\lambda _r f_{ri} } } \right) = \frac{1}
{{Z_q }}Nq_i \exp \left( { - \sum\limits_{r = 1}^R {\lambda _r f_{ri} } } \right), \qquad i = 1,...,s
\label{eq24}
\end{equation}
or
\begin{equation}
p_i ^*|q_i  = \frac{n_i ^*}{N} \approx q_i \exp \left( { - \lambda _0  - \sum\limits_{r=1}^R {\lambda _r f_{ri} } } \right) 
= \frac{1}{Z_q }q_i \exp \left( { - \sum\limits_{r=1}^R {\lambda _r f_{ri}}} \right), \qquad i = 1,...,s
\label{eq25}
\end{equation}
with
\begin{equation}
Z_q  = e^{\lambda _0 }  = \sum\limits_{i = 1}^s {q_i \exp \left( { - \sum\limits_{r = 1}^R {\lambda _r f_{ri} } } \right)} 
\label{eq26}
\end{equation}
where $p_i$ is the proportion or probability of entities in each state $i$.  Equations (\ref{eq24})-(\ref{eq25}) can be termed the generalized {\it Maxwell-Boltzmann distribution}, whilst $Z_q$ is the generalized {\it partition function} and $\lambda_0=\ln Z_q$ is the generalized {\it Massieu function} (strictly speaking, its negative \cite{Massieu_1869, Tribus_1961b}).  The Lagrangian multipliers are obtained from the constraints C{\it r} (\ref{eq17}) and/or more readily from moment calculations (see \S \ref{sect2_3_6}).

If ${\mathbf q} = {\mathbf u}$, (\ref{eq25}) reduces to:
\begin{equation}
\begin{split}
p_i ^*|u &\approx \frac{1}{Z}\exp \left( { - \sum\limits_{r = 1}^R {\lambda _r f_{ri} } } \right), \qquad i = 1,...,s  \\
& \qquad Z = \sum\limits_{i = 1}^s {\exp \left( { - \sum\limits_{r = 1}^R {\lambda _r f_{ri} } } \right)} 
\end{split}
\label{eq27}
\end{equation}
This is the more commonly reported, generalized {Maxwell-Boltzmann distribution} of statistical thermodynamics and information theory, and $Z$ is the usual generalized {partition function} \cite{Jaynes_1957, Tribus_1961b, Kapur_K_1992}.  Eq.\ (\ref{eq27}) is obtained directly if either $\ln \mathbb{P}_{\mathbf u}$ (\ref{eq14}) or $\ln \mathbb{W}$ (\ref{eq15}) is used in the Lagrangian (\ref{eq21}) instead of $\ln \mathbb{P}$.  

In the information literature, it is customary to cast the analysis in terms of $p_i$ rather than $n_i$, thus in ``$p_i$ form" \cite{Jaynes_1957, Tribus_1961a, Tribus_1961b, Kapur_K_1992}.  The constraints are:
\begin{alignat}{3}
&{\rm C0}:& \sum\limits_{i = 1}^s {p_i }&= 1&  \label{eq28} \\
&{\rm C1~to~C}R:& \qquad \sum\limits_{i = 1}^s {p_i f_{ri} }&= \left\langle {f_r } \right\rangle,&\qquad r = 1,...,R   \label{eq29}
\end{alignat}
hence the Lagrangian (\ref{eq21}) is:
\begin{equation}
\begin{split}
\mathfrak{L} &= \sum\limits_{i = 1}^s {\left( {p_i \ln N! - \ln[(p_i N)!] + p_i N\ln q_i } \right)}   \\*
&\qquad - (\mu _0^{}  - N)\left( {\sum\limits_{i = 1}^s {p_i }  - 1} \right) - \sum\limits_{r = 1}^R {\mu _r^{} \left( {\sum\limits_{i = 1}^s {p_i f_{ri} }  - \left\langle {f_r } \right\rangle } \right)} 
\end{split}
\label{eq30}
\end{equation}
where $\mu_r$, $r=0,...,R$, are the new Lagrangian multipliers, and $(\mu_0-N)$ is used for convenience.  Taking the variation and applying the Stirling approximation gives:
\begin{equation}
p_i ^*|q_i  \approx q_i \exp \left( { - \frac{{\mu _0^{} }}{N} - \sum\limits_{r = 1}^R {\frac{{\mu _r^{} }}
{N}f_{ri} } } \right) = \frac{1}{{Z_q '}}q_i \exp \left( { - \sum\limits_{r = 1}^R {\frac{{\mu _r^{} }}{N}f_{ri} } } \right), \qquad i = 1,...,s
\label{eq31}
\end{equation}
with	
\begin{equation}
Z_q' = e^{\mu _0 /N}  = \sum\limits_{i = 1}^s {q_i \exp \left( { - \sum\limits_{r = 1}^R {\frac{{\mu _r^{} }}
{N}f_{ri} } } \right)} 
\label{eq32}
\end{equation}
This is identical to (\ref{eq25})-(\ref{eq26}), with $\lambda_r = \mu_r/N$, $r=0,...,R$ and $Z_q' = Z_q$.  The Lagrangian multipliers are again obtained from the constraints (\ref{eq29}).

It is worth commenting that if the leading $\ln N!$ term is not brought inside the summation in (\ref{eq20}), but discarded - the approach of all previous workers - the resulting distribution $p_i^*$ contains an additional dependence on $N^{-1}$, which cancels out when forming the partition function $Z_q'' = Ne^{\lambda_0}$.  It therefore has no effect on traditional statistical mechanics.  The distinction is, however, important in the development of exact (finite-$N$) statistical mechanics, which does not use the Stirling approximation \cite{Niven_2005, Niven_2006}. 
Furthermore, by virtue of the natural constraint (\ref{eq16}), the $n_i$ form of a problem contains knowledge of $N$, and so it is unnecessary to include $N$ as a separate parameter. In contrast, a problem specified in its $p_i$ form does not contain - of itself - any information about $N$; if needed, this must be specified separately. 

From the foregoing it is clear that the ``most probable" probability distribution for a multinomial system, subject to arbitrary moment constraints, can be obtained {\it without making use of an entropy or cross-entropy function}.  One can instead analyse a probabilistic system directly using first combinatorial principles.  This aspect of entropy theory is not clearly spelt out in the information theory literature, with only a few exceptions \citep[e.g.][]{Jaynes_1963, Jaynes_1968, Jaynes_2003, Snickars_W_1977, Jefferson_S_1986, Kapur_1989a}. The direct combinatorial approach is extended further in \S \ref{sect3_2}, to encompass systems not of multinomial character.

\subsubsection{\label{sect2_3_3}Definition of the Cross-Entropy (Directed Divergence) and Entropy} 

Where do the cross-entropy and entropy functions come into the above analyses?  Clearly, they are merely convenient mathematical tools to enable construction of the Lagrangian equation in $p_i$ form (\ref{eq30}).  In fact we can {\it define} the cross-entropy as ``that function which, when inserted into the Lagrangian in place of $\ln \mathbb{P}$, and the extremum of the Lagrangian is obtained, yields the most probable distribution of the system".  The entropy may be similarly {\it defined} as ``that function which, when inserted into the Lagrangian in place of $\ln \mathbb{P}_{\mathbf u}$ (or $\ln \mathbb{W}$), and the extremum of the Lagrangian is obtained, yields the most probable distribution of the system."    

Consider $\ln \mathbb{P}$, expressed in $p_i$ form:
\begin{equation}
\ln \mathbb{P} = \sum\limits_{i = 1}^s {\left( {p_i \ln N! - \ln [(p_i N)!] + p_i N\ln q_i } \right)} 
\label{eq33}
\end{equation}
whence from the Stirling approximation (\ref{eq23}) \cite{Kapur_1989a}:
\begin{equation}
\ln \mathbb{P} \approx  - N\sum\limits_{i=1}^s {p_i \ln \frac{p_i}{q_i} } = -ND
\label{eq34}
\end{equation}
Thus the cross-entropy or directed divergence $D$ (\ref{eq6}) is simply the negative of the logarithm of the governing probability distribution, expressed per number of entities present \cite{Kapur_1989a}.  Maximizing $\ln \mathbb{P}$ for a multinomial system subject to the Stirling limits is therefore equivalent to maximizing $-D$, or minimizing $D$, subject to the constraints on the system.  (It does not matter whether we adopt a positive function, whose minimum yields the most probable realization, or its negative, whose maximum also yields this realization.  By convention, the cross-entropy is taken here as a positive function to be minimized, although this choice is arbitrary.)

Similarly if we consider $\ln \mathbb{P}_{\mathbf u}$, from (\ref{eq28}) and (\ref{eq34}) the Stirling form is \cite{Boltzmann_1877, Jaynes_1963, Kapur_1989a}:
\begin{equation}
\ln \mathbb{P}_{\mathbf{u}} \approx  - N\sum\limits_{i = 1}^s {p_i \ln sp_i }  
=  - N\ln s + NH
\label{eq35}
\end{equation}
This is proportional to the Shannon entropy (\ref{eq5}), shifted by a constant.  Maximizing $\ln \mathbb{P}_{\mathbf u}$ subject to the Stirling limits and constraints is therefore equivalent to maximizing $H$, subject to the same constraints \cite{Kapur_1989a}.  Indeed, from (\ref{eq15}):
\begin{equation}
\ln \mathbb{W} \approx  - N\sum\limits_{i = 1}^s {p_i \ln p_i }  = NH
\label{eq36}
\end{equation}
This definition of entropy for a multinomial system accords with the probabilistic expressions of Boltzmann and Shannon (\ref{eq5}).
  
It is therefore seen that the Kullback-Leibler cross-entropy and Shannon entropy functions are simplified forms of the logarithm of the multinomial distribution (\ref{eq13}), expressed per unit entity.  The MinXEnt and MaxEnt principles therefore provide simplified methods to determine the most probable realization of a multinomial system, subject to its constraints (succinctly termed ``MaxProb'' by Grendar and Grendar \cite{Grendar_G_2001}).  The cross-entropy is the more generic of the two functions, in that it contains the prior probabilities $q_i$.
    
Of the three theoretical roots of the entropy and cross-entropy functions, the combinatorial approach is therefore the most intellectually satisfying in that it provides a direct answer to the question: what is entropy?  There is no circular argument: entropy and cross-entropy are firmly based on simple combinatorial principles.  In consequence, there is no need to imbue either the MinXEnt or MaxEnt principles, or the cross-entropy or entropy functions themselves, with the kind of mystique with which they have been associated for well over a century.  There is no mystery at all.  In later sections, the foregoing analysis is generalized to any probabilistic system, irrespective of whether it is of multinomial character.

\subsubsection{\label{sect2_3_4}Equivalence of Reference States} 

It is necessary to be extremely careful about the definitions of the cross-entropy and entropy functions, given in \S \ref{sect2_3_3}.  To this end, note that obtaining the extremum of the Lagrangian ((\ref{eq21}) or (\ref{eq30})) necessitates extremization, whether it contains $\ln \mathbb{P}$ or its substitute, $-D$ (or whether $\ln \mathbb{W}$ or $H$, if ${\mathbf q}={\mathbf u}$).  The relationship between these quantities is therefore:
\begin{equation}
d\left( { - D({\mathbf{p}}|{\mathbf{q}})} \right) = \frac{1}{N}d\left( {\ln \mathbb{P}} \right)
\label{eq37}
\end{equation}
(In the present analysis, $\lambda_0,...,\lambda_R$ in the Lagrangian can be multiplied by any arbitrary positive constant $K$, and still give the same distribution, and so we could relax (\ref{eq37}) further by extremizing the scaled negative cross-entropy $-KD$.  This explains why we can use the scaled entropy $S=kH$ throughout thermodynamics, without affecting any calculations.)  Correspondence between the $i$th terms of $D$ and $\ln \mathbb{P}$ gives:
\begin{equation}
- \frac{\partial }{{\partial p_i }}D_i (p_i |q_i )dp_i  = \frac{1}{N} \frac{\partial }{{\partial p_i }}\ln \mathbb{P}_i dp_i \qquad i = 1,...,s
\label{eq38}
\end{equation}
where
\begin{equation}
D({\mathbf{p}}|{\mathbf{q}}) = \sum\limits_{i = 1}^s {D_i (p_i |q_i )} ,
\qquad \ln \mathbb{P} = \sum\limits_{i = 1}^s {\ln \mathbb{P}_i } \nonumber
\end{equation}
Integration with respect to $p_i$ and summation gives:
\begin{equation}
 - D({\mathbf{p}}|{\mathbf{q}}) = \frac{1}{N}\sum\limits_{i = 1}^s {\int {\frac{d}{{dp_i }}\ln \mathbb{P}_i dp_i } }  = \frac{{\ln \mathbb{P}}}
{N} + C
\label{eq39}
\end{equation}
where $C$ is a constant of integration.  In consequence, the multinomial cross-entropy (\ref{eq6}) and entropy (\ref{eq5}) could have been given respectively as (or any multiple of):
\begin{equation}
D({\mathbf{p}}|{\mathbf{q}}) = C + \sum\limits_{i = 1}^s {p_i \ln \frac{{p_i }}
{{q_i }}}  = \sum\limits_{i = 1}^s {\left( {Cp_i  + p_i \ln \frac{{p_i }}
{{q_i }}} \right)} 
\label{eq40}
\end{equation}
\begin{equation}
H({\mathbf{p}}) =  - C - \sum\limits_{i = 1}^s {p_i \ln p_i }  =  - \sum\limits_{i = 1}^s {\left( {Cp_i  + p_i \ln p_i } \right)} 
\label{eq41}
\end{equation}
However, the axiomatic definitions of these functions require that they obey the decisivity property (\S \ref{sect2_2}), i.e. $D=H=0$ when \{$p_i=1, i=j$; $p_i=0, i \ne j$\}, from which $C=0$, producing the recognized forms of the above functions ((\ref{eq6}), (\ref{eq5})).  This causes the $N \ln s$ term to be dropped from the definition of $H$ (\ref{eq35}).  Note, however, that the choice of $C$ has no impact on the application of $D$ or $H$ to determine the most probable realization.  (In other words, as is recognized throughout science and engineering, all zero reference or datum positions for the cross-entropy and entropy - and hence for information and energy - are mathematically equivalent.)  

\subsubsection{\label{sect2_3_5}Ensemble Theory and Multicomponent Systems} 

In its application to thermodynamics, one aspect of statistical mechanics has caused needless conceptual difficulty: the use of {\it ensembles} to represent particular types of systems \cite{Gibbs_1902}.  Most common are the {\it microcanonical ensemble}, representing a closed system of fixed energy; the {\it canonical ensemble}, a closed system of fixed temperature; and the {\it grand canonical ensemble}, an open system of fixed temperature and mean composition.  From the foregoing discussion, it is evident that an ensemble is simply {\it the set of all possible realizations - each weighted by its number of permutations (or for unequal} $q_i$, {\it by the probability of each realization) - consistent with a particular system specification}; i.e. consistent with a specified governing probability distribution $\mathbb{P}$, total number of entities $N$ (or numbers of entities of different types), number of states $s$, and specified constraints $\left\langle {f_r} \right\rangle$ or their equivalent Lagrangian multipliers $\lambda_r, r=1,...,R$.  An ensemble is therefore a {\it mental} construct, which does not require a physical manifestation.  

As an example, consider a closed physical system in which the entities fluctuate between states (the {\it elemental chaos} of Planck \cite{Planck_1913}), such as gas molecules in a container.  Such a system will migrate from one realization to another, and thence between different members of its ensemble (it will describe a trajectory in - for example - energetic, geometric or phase space).  
However, there is no need to require that the system {\it must} access every realization within a particular time frame, nor even that it should come arbitrary close to every realization; the only requirement of probability theory is that each realization included in the ensemble be realizable, to the extent given by its assigned probability.  As every gambler or insurance broker will testify, probabilities are not certainties.  Unfortunately, a great deal of erroneous reasoning has been put forth on this topic, which still clouds our present-day understanding.

In contrast, consider a ``multiple selection" system as defined in \S \ref{sect2_3_1}, such as a set of throws of a coin or rolls of a die.  In this case, the ensemble can only ever be a mental construct, representing the set of all possible outcomes.  Once the ``die is cast", the ensemble ceases to have any meaning, except as a reminder of what ``might have been".     

The microcanonical and canonical ensembles are both based on the multinomial distribution (\ref{eq13}), with different interpretations.  In the (generalized) microcanonical ensemble, $N$ represents the total number of non-interacting particles, each of which is deemed to possess its own ``private" functions $f_{ri}$. The constraints $\left\langle {f_r} \right\rangle$ can therefore be considered constant.  In contrast, in the (generalized) canonical ensemble, $N$ is now the number of separate systems (this is more clearly denoted $\mathbb{N}$ \cite{Atkins_1982}), each of which contains a constant number of particles, all subject to baths of constant $\lambda_r, r=1,...,R$.  By this device, the canonical ensemble can be used to examine systems containing interacting particles\ 
\!\footnote{The precise definition of ``interacting" remains open.  Some workers prefer to qualify this statement, by considering only ``weakly interacting" particles \citep[e.g.][]{Jaynes_1957, Eyring_1964} or those without ``long-range interactions" \citep[e.g.][]{Tsallis_2001}.}
or other coupling effects, thus in which the $f_{ri}$ (hence the $\left\langle {f_r} \right\rangle$) can be functions of the realization, even though the $\lambda_r$ are fixed \citep[see][]{Gibbs_1902, Einstein_1902, Einstein_1903, Schrodinger_1952}.  In other words, the canonical ensemble represents ``the set of realizations of the set of realizations of interacting particles."  This superset cannot readily be reduced to the lower (microcanonical) set unless the particles are non-interacting.  Despite this distinction, by the use of baths of ``generalized heat" (see \S \ref{sect2_3_6}), the canonical ensemble is analysed by the same mathematical treatment as the microcanonical \cite{Schrodinger_1952, Jaynes_1957}.

The generalized grand canonical ensemble is normally taken to consist of $\mathbb{N}$ separate systems, in which there are $n_{\{ N_l \} ,i}$ systems containing $N_l$ particles each of the $l$th type in the $i$th state, for $l=1,...,L$, where $L$ is the number of independent species.  (For reactive species, it is necessary to define a minimum set of $L$ species, from which all other species can be formed by reaction \cite{Gibbs_1875}.)  Since the system is open, each $N_l$ is permitted to vary between zero and (effectively) infinity.  Expressed in terms of $\mathbb{P}$ rather than $\mathbb{W}$, the governing distribution is generally assumed to be ``multiply multinomial" \citep[c.f.][]{Hill_1956, Desloge_1966, Abe_1975, McQuarrie_1976}:
\begin{equation}
\mathbb{P}_{GC}  = \mathbb{N}! \prod\limits_{N_1  = 0}^\infty  ...\prod\limits_{N_L  = 0}^\infty  
{\prod\limits_{i=1}^s {\frac{{q_{\{ N_l \} ,i} ^{\hspace{20pt} n_{\{ N_l \} ,i} } }} {{n_{\{ N_l \} ,i} !}}} } 
\label{eq42}
\end{equation}
where $q_{\{ N_l \} ,i}$ is the prior probability of a system which contains $N_l$ particles of each $l$th type in the $i$th state.  This is normally subject to natural, moment and mean number of each type of entity constraints:
\begin{alignat}{3}
&{\rm C}0: &\qquad \sum\limits_{N_1  = 0}^\infty  {...\sum\limits_{N_L  = 0}^\infty  {\sum\limits_{i = 1}^s {n_{\{ N_l \} ,i} } } }  &= \mathbb{N} \qquad \label{eq43} \\
&{\rm C}r:& \qquad \sum\limits_{N_1  = 0}^\infty  {...\sum\limits_{N_L  = 0}^\infty  {\sum\limits_{i = 1}^s {n_{\{ N_l \} ,i} f_{ri} } } }  &= \mathbb{N}\left\langle {f_r } \right\rangle, &\qquad r = 1,...,R  
\label{eq44} \\
&{\rm C} l:& \qquad \sum\limits_{N_1  = 0}^\infty  {...\sum\limits_{N_L  = 0}^\infty  {\sum\limits_{i = 1}^s {n_{\{ N_l \} ,i} N_l } } }  &= \mathbb{N}\left\langle {N_l } \right\rangle, &\qquad l = 1,...,L
\label{eq45}
\end{alignat}
The combinatorial method ((\ref{eq39}) and \S \ref{sect2_3_2}) gives the Stirling-approximate cross-entropy and equilibrium distribution:
\begin{gather}
- D_{GC}  = \frac{{\ln \mathbb{P}}}
{\mathbb{N}} \approx  - \sum\limits_{N_1  = 0}^\infty  {...\sum\limits_{N_L  = 0}^\infty  {\sum\limits_{i = 1}^s {p_{\{ N_l \} ,i} } } } \ln \frac{{p_{\{ N_l \} ,i} }}
{{q_{\{ N_l \} ,i} }}
\label{eq46} \\
p_{\{ N_l \} ,i} ^* = \frac{1}
{{\Xi _q }}q_{\{ N_l \} ,i} \exp \left( { - \sum\limits_{r = 1}^R {\lambda _r f_{ri} }  - \sum\limits_{l = 1}^L {\nu _l N_l } } \right), \qquad i = 1,...,s
\label{eq47}
\end{gather}
with
\begin{equation}
\Xi _q  = \sum\limits_{N_1  = 0}^\infty  {...\sum\limits_{N_L  = 0}^\infty  {\sum\limits_{i = 1}^s {q_{\{ N_l \} ,i} \exp \left( { - \sum\limits_{r = 1}^R {\lambda _r f_{ri} }  - \sum\limits_{l = 1}^L {\nu _l N_l } } \right)} } } 
\label{eq48}
\end{equation}
where $p_{\{ N_l \} ,i}  = n_{\{ N_l \} ,i} /\mathbb{N}$; $\lambda_r$ and $\nu_l$ are Lagrangian multipliers; and $\Xi_q$ is the generalized grand partition function.  The entropy forms follow.  However, the cross-entropy will only be of Kullback-Leibler form if the governing distribution is multinomial (\ref{eq42}).  If $\mathbb{P}$ is of some other form, for example the product of independent distributions (extending \cite{Davidson_1962}):
\begin{equation}
\mathbb{P}_{GC} \hspace{2pt}' = \prod\limits_{l = 0}^L {\mathbb{P}_l  = } \prod\limits_{l = 0}^L {\prod\limits_{N_l  = 0}^\infty  {\mathbb{N}_l ! \prod\limits_{i = 1}^s {\frac{{q_{N_l ,i} ^{\hspace{10pt} n_{N_l ,i} } }}{{n_{N_l ,i} !}}} } },
\qquad
\sum\limits_{N_l  = 0}^\infty  {\sum\limits_{i = 1}^s {n_{N_l ,i} } }  = \mathbb{N}_l 
\label{eq49}
\end{equation}
or if we possess some other knowledge (such as of $N_l$), then clearly the resulting multicomponent cross-entropy and entropy functions and the equilibrium distribution could be quite different.  It is insufficient to simply assert (\ref{eq42}) or (\ref{eq46}); its adoption must be based on sound reasoning, and ultimately, be demonstrated by successful predictions.

\subsubsection{\label{sect2_3_6}``Jaynes Relations" and Generalized Free Energy Function} 

It is now possible to re-examine the generic structure of statistical mechanics developed by Jaynes \citep[e.g.][]{Jaynes_1957, Jaynes_1963, Jaynes_1968, Jaynes_1978, Jaynes_2003, Tribus_1961a, Tribus_1961b, Kapur_K_1987, Kapur_K_1992}, in light of the combinatorial approach. The generic structure has the powerful advantage of being applicable to {\it any} multinomial system, irrespective of the physical nature of the constraints, and is therefore not limited to energetic systems.  It can however be even further extended. The following discussion is a synthesis and extension of previous treatments, building up to the development of major new concepts including a generalized Clausius inequality, a generalized free energy function, and a generalized Gibbs-Duhem relation and phase rule.  Throughout the following (except where specified), $\lambda_0=\ln Z_q$ is assumed to be a function of each $\lambda_r$; the $\lambda_r$ are mutually independent; each $f_{ri}$ is independent of $\lambda_r$; and each $\left\langle {f_r} \right\rangle$ is a function of $\lambda_r$ but not of the other multipliers $\lambda_m$, $m \ne r$.  

From $p_i^*$ (\ref{eq24})-(\ref{eq25}) and the moment constraints (\ref{eq17}) it can be shown that \cite{Jaynes_1957, Tribus_1961b, Kapur_K_1992}:
\begin{equation}
- \frac{{\partial \lambda _0 }}{{\partial \lambda _r }} = \left\langle {f_r } \right\rangle 
\label{eq50}
\end{equation}
The variance and covariances of $f_{ri}$, necessarily in the vicinity of equilibrium, are obtained by further differentiation \cite{Jaynes_1957, Jaynes_1963, Tribus_1961a, Tribus_1961b, Kapur_K_1992}:
\begin{gather}
\frac{{\partial ^2 \lambda _0 }}
{{\partial \lambda _r ^2 }} = {\rm var} (f_r ) = \left\langle {f_r ^2 } \right\rangle  - \left\langle {f_r } \right\rangle ^2  =  - \frac{{\partial \left\langle {f_r } \right\rangle }}{{\partial \lambda _r }}
\label{eq51} \\
\frac{{\partial ^2 \lambda _0 }}
{{\partial \lambda _m \partial \lambda _r }} = {\rm cov} (f_m ,f_r ) = \left\langle {f_r f_m } \right\rangle  - \left\langle {f_r } \right\rangle \left\langle {f_m } \right\rangle  =  - \frac{{\partial \left\langle {f_r } \right\rangle }}{{\partial \lambda _m }}
\label{eq52} 
\end{gather}
where each $f_{ri}$ is independent of each $\lambda_m$.  From (\ref{eq52}), ${\partial ^2 \lambda_0} / {\partial \lambda_m \partial \lambda_r} = {\partial ^2 \lambda_0 } / {\partial \lambda_r \partial \lambda_m}$, whence the coupling coefficients are equal:
\begin{equation}
\frac{{\partial \left\langle {f_r } \right\rangle }}
{{\partial \lambda _m }} = \frac{{\partial \left\langle {f_m } \right\rangle }}
{{\partial \lambda _r }}
\label{eq53}
\end{equation}
Eq.\ (\ref{eq52}) is a subset of a more general result \cite{Jaynes_2003}:
\begin{equation}
{\rm cov} (g,f_r ) = \left\langle {gf_r } \right\rangle  - \left\langle g \right\rangle \left\langle {f_r } \right\rangle  =  - \frac{{\partial \left\langle g \right\rangle }}
{{\partial \lambda _r }}
\label{eq54}
\end{equation}
where ${\mathbf g}=\{ g_i \}$ is any function of the states $i=1,...,s$, in which each $g_i$ is independent of $\lambda_r$.  

Using the Cauchy-Schwartz inequality $\left\langle {a^2} \right\rangle \left\langle {b^2 } \right\rangle  - \left\langle {ab} \right\rangle ^2  \geq 0$ \cite{Zwillinger_2003} with $a=f_r, b=1$ gives ${\rm var} (f_r) = - \partial \left\langle {f_r} \right\rangle /\partial \lambda_r  \geq 0$, whence $\partial \left\langle {f_r} \right\rangle /\partial \lambda _r \leq 0$ \cite{Kapur_K_1987}.  Accordingly, $\lambda_r$ decreases monotonically with increasing $\left\langle {f_r} \right\rangle$.  No equivalent relation is available for the mixed derivatives $\partial \left\langle {f_r} \right\rangle /\partial \lambda_m$.  Using the arguments of Kapur and Kesevan \cite[\S2.4.2; 4.3.2]{Kapur_K_1992}, we find that $\lambda_0$ is a convex function of $\lambda_r$, $r=1,...,R$.

It is also possible to consider $\lambda_0$ and each $f_{ri}$ (hence also $\left\langle {f_r} \right\rangle$) to be functions of parameters $\alpha_v$, $v=1,...,V$.  By differentiation of the partition function (\ref{eq26}) \cite{Jaynes_1957, Jaynes_1963, Jaynes_2003}, or more directly by rearrangement of $p_i^*$ ((\ref{eq24})-(\ref{eq25})) and differentiation:
\begin{equation}
- \frac{{\partial \lambda _0 }}{{\partial \alpha _v }} = \sum\limits_{r = 1}^R {\lambda _r } \left\langle {\frac{{\partial f_r }}{{\partial \alpha _v }}} \right\rangle , \qquad v = 1,...,V
\label{eq55}
\end{equation}
Alternatively, differentiation of (\ref{eq53}) with respect to any continuous function $\alpha_v$ yields (necessarily in the vicinity of equilibrium, e.g.\  for a shifting equilibrium position):
\begin{equation}
\frac{\partial }{{\partial \lambda _m }}\left( {\frac{{\partial \left\langle {f_r } \right\rangle }}{{\partial \alpha _v }}} \right) = \frac{\partial } {{\partial \lambda _r }}\left( {\frac{{\partial \left\langle {f_m } \right\rangle }}{{\partial \alpha _v }}} \right)
\label{eq56}
\end{equation}
Eq.\ (\ref{eq56}) with $\alpha_v=t=$ time is a statement of Onsager's \cite{Onsager_1931a, Onsager_1931b} reciprocal relations.  Various other higher derivative equations in $\lambda_r$ and/or $\alpha_v$ are given by Jaynes \cite{Jaynes_2003}.

Similarly, considering $\lambda_0$ and $\lambda_r$ to be functions of $\beta_j$, $j=1,...,J$; or $\lambda_0$ alone as a function of $N$, $n_i^*$ or $p_i^*$, from (\ref{eq24})-(\ref{eq26}):
\begin{alignat}{2}
- \frac{{\partial \lambda _0 }}{{\partial \beta _j }} &= \sum\limits_{r = 1}^R {\frac{{\partial \lambda _r }}
{{\partial \beta _j }}} \left\langle {f_r } \right\rangle , &\qquad j = 1,...,J
\label{eq57} \\
\frac{{\partial \lambda _0 }}{{\partial N}} &= 0
\label{eq58} \\
- \frac{\partial \lambda _0 } {\partial n_i ^*} &= \frac{1}{n_i ^*}, &\qquad
- \left\langle \frac{\partial \lambda _0 }{\partial n_i ^*} \right\rangle  &= \left\langle \frac{1}{n_i ^*} \right\rangle  = \frac{s} {N} 
\label{eq59} \\
- \frac{{\partial \lambda _0 }}{{\partial p_i ^*}} &= \frac{1}{{p_i ^*}}, &\qquad
- \left\langle {\frac{{\partial \lambda _0 }}{{\partial p_i ^*}}} \right\rangle  &= \left\langle {\frac{1}{{p_i ^*}}} \right\rangle  = s 
\label{eq60}
\end{alignat}
From (\ref{eq58}), $\lambda_0$ (and thus $Z_q$) is independent of $N$ in the Stirling limit $N \to \infty $.  From (\ref{eq59}), $\left\langle {\partial \lambda _0 /\partial n_i ^*} \right\rangle  \to 0$ in the Stirling limit $n_i^* \to \infty$, hence $\lambda_0$ is independent of the mean degree of filling of each state.

Using $p_i^*$ ((\ref{eq24})-(\ref{eq25})), the constraints ((\ref{eq16})-(\ref{eq17}) or (\ref{eq28})-(\ref{eq29})), the definitions of $H$, $D$ and $\mathbb{P}$ ((\ref{eq5})-(\ref{eq6}),(\ref{eq39})) and the multiplier relations ((\ref{eq50})), the minimum cross-entropy or maximum entropy position is obtained as \citep[c.f.][]{Jaynes_1957, Tribus_1961b, Kapur_K_1992}:
\begin{equation}
- D^* = H^* = \lambda _0  + \sum\limits_{r = 1}^R {\lambda _r \left\langle {f_r } \right\rangle }  = \ln Z_q  - \sum\limits_{r = 1}^R {\lambda _r \frac{{\partial \ln Z_q }}{{\partial \lambda _r }}} 
\label{eq61}
\end{equation}
with probability:
\begin{equation}
\mathbb{P}^* = A\exp ( - ND^*)
\label{eq62}
\end{equation}	
where $A$ is a normalising constant (with $\mathbb{P}^* \leq 1$), and we recall that $H^*$ is obtained from $\ln \mathbb{P}_{\mathbf u}$ by dropping the $\ln s$ term (or directly from $\ln \mathbb{W}$) ((\ref{eq35})-(\ref{eq36})).  Equation (\ref{eq61}) is one of the most important equations in equilibrium statistical mechanics - for example giving the thermodynamic entropy and thence all thermodynamic functions in terms of the applicable partition function - whilst (\ref{eq62}) encompasses Einstein's \cite{Einstein_1904} definition of entropy.  Note that the MinXEnt and MaxEnt positions are of the same form, although ${\mathbf q}$ is implicit within $\lambda_0$ in $D^*$.  By successive differentiation of (\ref{eq61}) with respect to the moments - taking $\lambda_0$ to be independent of $\left\langle {f_r} \right\rangle$ - gives \citep[c.f.][]{Jaynes_1957, Jaynes_1963, Jaynes_2003, Kapur_K_1992}:
\begin{gather}
- \frac{{\partial D^*}}{{\partial \left\langle {f_r } \right\rangle }} = \frac{{\partial H^*}}{{\partial \left\langle {f_r } \right\rangle }} = \lambda _r 
\label{eq63} \\
- \frac{{\partial ^2 D^*}}
{{\partial \left\langle {f_m } \right\rangle \partial \left\langle {f_r } \right\rangle }} = \frac{{\partial ^2 H^*}}
{{\partial \left\langle {f_m } \right\rangle \partial \left\langle {f_r } \right\rangle }} = \frac{{\partial \lambda _r }}
{{\partial \left\langle {f_m } \right\rangle }} = \frac{{\partial \lambda _m }}
{{\partial \left\langle {f_r } \right\rangle }}
\label{eq64}
\end{gather}
whilst differentiation with respect to $\lambda_r$ - now considering $\left\langle {f_r } \right\rangle$ to be a function of $\lambda_m, \forall m$ - and use of (\ref{eq53}) gives the Euler relation \citep[c.f.][]{Plastino_1997}:
\begin{equation}
 - \frac{{\partial D^*}}
{{\partial \lambda _r }} = \frac{{\partial H^*}}
{{\partial \lambda _r }} = \sum\limits_{m = 1}^M {\lambda _m \frac{{\partial \left\langle {f_m } \right\rangle }}
{{\partial \lambda _r }}}  = \sum\limits_{m = 1}^M {\lambda _m \frac{{\partial \left\langle {f_r } \right\rangle }}
{{\partial \lambda _m }}} 
\label{eq65}
\end{equation}
where $M$ and $R$ are numerically equal.  From (\ref{eq63}), using the same arguments as Kapur \& Kesevan \cite[\S2.4.4; 4.3.2]{Kapur_K_1992}, we see that $D^*$ (or $H^*$) is a convex (concave) function of the $\left\langle {f_r } \right\rangle$'s.  A multinomial system subject to the Stirling approximation therefore has a single, unique equilibrium position with respect to its moment constraints.

The variation in $D^*$ or $H^*$ due to variations in $\lambda_0$, $\lambda_r$ and $\left\langle {f_r} \right\rangle$ (and also $N$) is \citep[c.f.][]{Jaynes_1957, Jaynes_1963, Jaynes_2003, Tribus_1961b}:
\begin{equation}
-dD^* = dH^* = \sum\limits_{r = 1}^R {\lambda _r (d\left\langle {f_r } \right\rangle  - \left\langle {df_r } \right\rangle )}  = \sum\limits_{r = 1}^R {\lambda _r dQ_r } 
\label{eq66}
\end{equation}
where we can interpret $d\left\langle {f_r} \right\rangle = dU_r$, $\left\langle {df_r} \right\rangle = \sum\nolimits_{i = 1}^s {p_i df_{ri} } = dW_r$ and $d\left\langle {f_r } \right\rangle  - \left\langle {df_r } \right\rangle  = \sum\nolimits_{i = 1}^s {f_{ri} } dp_i = dQ_r$ respectively as changes in the $r$th type of ``energy", ``generalized work" on the system and ``generalized heat" delivered to the system
, whence (as defined here) $dU_r = dQ_r + dW_r$.  Note that in the above derivation, the variations in $\lambda_r$ cancel out \cite{Jaynes_1963, Jaynes_2003}, hence (\ref{eq66}) encompasses conditions of either constant or variable $\lambda_r$.  Equation (\ref{eq66}) is a superset of the Clausius relation (\ref{eq1}), and so for each type of ``generalized heat" there exists a conjugate integrating factor $\lambda_r$.  As with the Clausius relation, the $\lambda_r$ are properties of the system of interest (i.e.\ the one into which positive generalized heat is delivered).

Equation (\ref{eq66}) applies to a reversible process, i.e.\ to an incremental change in the equilibrium position.  If we also include spontaneous irreversible processes (involving a system not necessarily at equilibrium), for which the cross-entropy can decrease (or entropy can increase) without generalized heat input, we see that:
\begin{equation}
-dD = dH \geq \sum\limits_{r = 1}^R {\lambda _r dQ_r } 
\label{eq67}
\end{equation}
This is a superset of the Clausius inequality (\ref{eq2}).  Equation (\ref{eq67}) can be rearranged, in the manner of Gibbs \cite{Gibbs_1875, Gibbs_1902}, to give the differential form of a generic dimensionless free energy function $\Phi$, here termed the {\it free information}\
\!\footnote{This is quite distinct from the ``free physical information'' of Frieden \cite{Frieden_2004}.}:
\begin{equation}
d\Phi  = \left\{ {\begin{array}{*{20}c}
   {dD + \sum\limits_{r = 1}^R {\lambda _r dQ_r } }  \\
   { - dH + \sum\limits_{r = 1}^R {\lambda _r dQ_r } }  \\
\end{array} } \right\} \leq 0
\label{eq68}
\end{equation}
(whence $d\Phi ^* = 0$ at a fixed equilibrium position), where the upper form incorporates the prior probabilities ${\mathbf q}$.  Now from (\ref{eq61}):
\begin{equation}
-dD^* = dH^* = d\lambda _0  + \sum\limits_{r = 1}^R {d\lambda _r \left\langle {f_r } \right\rangle }  + \sum\limits_{r = 1}^R {\lambda _r d\left\langle {f_r } \right\rangle } 
\label{eq69}
\end{equation}
so if we set $dD = dD^* + dD^{irrev}$ and $dH = dH^* + dH^{irrev}$ (with $dD^{irrev} \leq 0$ and $dH^{irrev}  \geq 0$), where superscript {\it irrev} denotes the irreversible component, then from (\ref{eq68})-(\ref{eq69}):
\begin{equation}
d\Phi  = \left\{ {\begin{array}{*{20}c}
   { - d\lambda _0  - \sum\limits_{r = 1}^R {d\lambda _r \left\langle {f_r } \right\rangle }  + dD^{irrev}  - \sum\limits_{r = 1}^R {\lambda _r dW_r } }  \\
   { - d\lambda _0  - \sum\limits_{r = 1}^R {d\lambda _r \left\langle {f_r } \right\rangle }  - dH^{irrev}  - \sum\limits_{r = 1}^R {\lambda _r dW_r } }  \\
\end{array} } \right\} \leq 0
\label{eq70}
\end{equation}
If - and only if - there is no change in $\lambda_r$ (i.e.\ no change in any contacting bath; see also (\ref{eq74}) below), no reversible generalized work on the system (apart from that already included in the constraints) and no irreversible process, then:
\begin{equation}
d\Phi ^* =  - d\lambda _0  =  - d\ln Z_q 
\label{eq71}
\end{equation}
where $Z_q$ is the applicable partition function ((\ref{eq26}) or (\ref{eq27})).  Alternatively, from (\ref{eq70}), if there is no change in $\lambda_0$ or $\lambda_r$ and no irreversible process:
\begin{equation}
d\Phi  =  - \sum\limits_{r = 1}^R {\lambda _r dW_r }  \leq 0
\label{eq72}
\end{equation}
$\Phi$ therefore indicates the maximum available weighted generalized work per entity which can be obtained from a system.  

Integration of (\ref{eq68}) gives the state function:
\begin{equation}
\Phi  = \left\{ {\begin{array}{*{20}c}
   {D + \sum\limits_{r = 1}^R {\lambda _r Q_r } }  \\
   { - H + \sum\limits_{r = 1}^R {\lambda _r Q_r } }  \\
\end{array} } \right.
\label{eq73}
\end{equation}
where $Q_r  = \int {dQ_r}  = \int d \left\langle {f_r} \right\rangle  - \int {dW_r}$ defines each absolute generalized heat\ 
\!\footnote{In thermodynamic systems, this is generally approximated as $Q_r  \approx \left\langle {f_r} \right\rangle$, i.e. assuming each generalized work term is zero, except for the energy constraint, where the actual heat $Q = \int {dQ}  = \int {TdS}  = TS$ at constant $T$ is used.}. Comparing its differential with (\ref{eq68}) gives:
\begin{equation}
\sum\limits_{r = 1}^R {Q_r d\lambda _r }  = 0
\label{eq74}
\end{equation}
This is a superset of the Gibbs-Duhem equation \cite{Gibbs_1875}.  For a system containing separate coexistent {\it phases}, or bodies which differ in composition or state (as defined by Gibbs \cite{Gibbs_1875}), there will be one such equation for each phase.  For $L$ independent constituents, $\mathfrak{r} = R - L$
other constraints (not including the $L$ constituents) and $\mathfrak{p}$ phases, (\ref{eq74}) thus yields a generalized Gibbs' phase rule for the number of degrees of freedom of a system \citep[c.f.][]{Gibbs_1875, Tribus_1961b, Eyring_1964}:
\begin{equation}
f = L + \mathfrak{r} - \mathfrak{p} = R - \mathfrak{p}
\label{eq75}
\end{equation}
In other words, the system will be fully determined by $R - \mathfrak{p}$ independent parameters, from the set of $R$ constraints or (more commonly) their corresponding Lagrangian multipliers.

Equations (\ref{eq61}), (\ref{eq68}) and (\ref{eq70})-(\ref{eq75}) form the basis of present-day thermodynamics.  For energetic systems, $d \Phi$ is normally divided by the energetic multiplier $\lambda_1 = 1/kT$; e.g.\  for an energetic system which can exchange heat with its surroundings, but not work or mass, at constant volume, $dQ_1  = dU$, $dS = kdH$, $dA = kTd\Phi  = dU - TdS \leq 0$ and $dA^*=-kTd\ln Z$, where $U$ is the mean internal energy per entity, $A$ is the Helmholtz free energy per entity and $Z$ is the microcanonical or canonical partition function\ 
\!\footnote{The extensive thermodynamic variables (e.g.\  $U, S, V, m_l, A, G$) are all mean quantities, expressed in relevant units per entity.  In a microcanonical ensemble, they represent mean values per particle.  The total values are calculated by multiplication by $N$ (the form of (\ref{eq68}) remains the same).  In a canonical ensemble, each extensive variable represents the ``ensemble mean" or ``mean of the total values".}
.  For a grand canonical system with $L$ independent constituents which can exchange heat and mass with its surroundings, but not work except for $PV$-work, at constant pressure, $dQ_1  = dU$, $\lambda _1  = 1/kT$, $dQ_2  = dV$, $\lambda _2  = P/kT$, $dQ_{2 + l}  = dm_l$, $\lambda _{2 + l}  =  - \mu _l /kT =  - \ln \alpha _l $, $dG = kTd\Phi  = dU - TdS + PdV - \sum\nolimits_l^{} {\mu _l dm_l }  \leq 0$, $dG^* =  - kTd\ln \Xi$
and $f = L + 2 - \mathfrak{p}$, where $P$ is pressure, $V$ is mean volume per entity, $\mu_l$ is the chemical potential and $\alpha_l$ is the ``absolute" (unscaled) chemical activity of the $l$th constituent, $m_l$ is the mean number of entities of $l$th type per entity, $G$ is the Gibbs free energy per entity and $\Xi$ is the grand canonical partition function. The {\it essergy} $Y = kT_0 \Phi  = E - T_0 S + P_0 V - \sum\nolimits_l^{} {\mu _{l0} m_l } $ is a scaled $\Phi$ of a system with total internal energy $E$, in contact with a bath of reference temperature $T_0$, pressure $P_0$ and chemical potentials $\{\mu_{l0}\}$ \cite{Evans_1969}.  Essergy is thus an extended free energy calculated with reference to the bath (e.g.\  the external environment), not to the system. The {\it exergy} $X=Y-Y_0$ is the difference between the essergy of a system (by early authors, with the chemical potential terms omitted), and of the same system in equilibrium with the bath \citep[e.g.][]{Keenan_1941, Keenan_1951, Rant_1956, Gaggioli_1962, Evans_1969, Ahern_1980, Scubbia_2005, Sciubba_U_2005}.  Exergy therefore represents the maximum work deliverable to the environment, by allowing a system to reach equilibrium with that environment.  The statistical {\it extropy} \cite {Martinas_1998, Martinas_2000, Gaveau_2002} is a modified free information defined with respect to the bath - with all generalized work terms set to zero (i.e.\ $Q_r  \approx \left\langle {f_r} \right\rangle$) - less the modified free information at equilibrium.  Exergy forms the nucleus of the interrelated fields of thermoeconomics and exergo-economics for resource management and process optimization \cite{Tribus_E_1962, Scubbia_2005, Valero_etal_2006}, whilst both exergy and extropy have been used as measures of environmental impact, i.e.\ as quantitative tools within and/or complementary to the framework of environmental life cycle assessment \cite{Ayres_etal_1998, Ayres_1998, Sciubba_U_2005, Martinas_1998, Martinas_2000}.

Notwithstanding the historical development of this field, it must be emphasized that the use of $\Phi$ is not restricted to thermodynamic, industrial or environmental systems.  Just as with the information entropy, we can define the free information of any multinomial system - for example in communications, transport, urban planning, biology, geography, social science, politics, economics, linguistics, image analysis or any other field - and use it to examine its (probabilistic) stability.  The entire armoury of state functions, cyclic integrals, efficiency ratios, Gibbs-Duhem and phase relations, Maxwell-like relations and Jaynes relations - currently considered the exclusive domain of thermodynamics - can then be brought to bear to the analysis of such systems.

\subsubsection{\label{sect2_3_7}``Fluctuations" and Entropy Concentration Theorem} 

Although the MinXEnt or MaxEnt distribution is the ``most probable" one, it cannot be {\it a priori} assumed to be the exclusive outcome.  The sharpness of the predicted distribution has historically been examined by two methods: the fluctuation criterion of Gibbs \cite{Gibbs_1902} and Einstein \cite{Einstein_1904}, and the entropy concentration theorem of Jaynes \cite{Jaynes_1968, Jaynes_1978, Jaynes_1979, Jaynes_2003}, in part foreshadowed by Boltzmann \cite{Boltzmann_1896} and Einstein \cite{Einstein_1905}.  The detailed asymptotic convergence behaviour of the distribution forms the subject of large deviations theory, based on various mathematical limit theorems \cite{Book_1983, Ellis_1985, Cover_T_1991}, and will not be examined further here.

The first method examines the coefficient of variation $\delta$ of each constraining variable (or its square), commonly termed its ``fluctuation"\ 
\!\footnote{The term ``fluctuation" is unfortunate, since it implies rapid change about the mean, which has little to do with the equilibrium position but depends on the system dynamics. $\delta (Nf_r)$ is simply a measure of the ``variability" or ``spread" of the equilibrium filling of $N\{f_{ri} \}$.}
. For a microcanonical system, this can be written as \citep[c.f.][]{Gibbs_1902, Einstein_1904}:
\begin{equation}
\delta (Nf_r ) = \frac{{\sqrt {{\rm var} (Nf_r )} }}
{{\left\langle {Nf_r } \right\rangle }} = \frac{{\sqrt {N\left[ {\left\langle {f_r ^2 } \right\rangle  - \left\langle {f_r } \right\rangle ^2 } \right]} }}
{{\left\langle {Nf_r } \right\rangle }}
\label{eq76}
\end{equation}
where we are careful with notation to consider the variability about the total extensive quantity $\left\langle {Nf_r } \right\rangle$ for a system of $N$ entities, not the variability of the fixed quantity per entity $\left\langle {f_r } \right\rangle $. (Of course, $\delta$ does not capture the full picture of the distribution of $N\{f_{ri} \}$, e.g.\  the skewness, kurtosis, etc, for which higher order moments must be considered.)  The criterion for sharpness is normally stated as $\delta \ll 1$ \cite{Einstein_1904, Jaynes_1963}.  From  (\ref{eq51}) and (\ref{eq76}):
\begin{equation}
\delta (Nf_r ) = \frac{1}
{{\sqrt N }}\sqrt { - \frac{1}
{{\left\langle {f_r } \right\rangle ^2 }}\frac{{\partial \left\langle {f_r } \right\rangle }}
{{\partial \lambda _r }}} 
\label{eq77}
\end{equation}
The term inside the second square root is positive, and in many cases of order unity, whereupon $\delta (Nf_r ) \approx N^{ - 1/2} \to 0$ in the Stirling limit $N \to \infty $. For example, for a microcanonical system with $f_{1i}  = \varepsilon _i $, $\left\langle {f_1 } \right\rangle  = \left\langle \varepsilon  \right\rangle  = U$, $\lambda _1  = 1/kT$, containing an ideal monatomic non-interacting gas with $U=\frac{3}{2}kT$ and $C_v  = \partial U/\partial T = \frac{3}{2}k$, where $C_v$ is the isovolumetric heat capacity per entity, we obtain $\delta (N\varepsilon ) = (\frac{3}{2}N)^{ - 1/2}  \approx N^{ - 1/2}  \to 0$ \citep[e.g.][]{Tolman_1938, Schrodinger_1952, Hill_1956, Davidson_1962, Eyring_1964, Desloge_1966, McQuarrie_1976}\ 
\!\footnote{\label{footcanon} All the listed authors consider $\delta (E)$ for a canonical ensemble, where $\left\langle E \right\rangle $ is the ``mean of the total energies", but then take $\left\langle E \right\rangle =N\left\langle \varepsilon  \right\rangle = \frac{3}{2} NkT$ for $N$ non-interacting particles - thus assuming the system is microcanonical - giving the same result.}\ 
.  Although this result is not general (e.g.\  in the vicinity of phase changes \cite{McQuarrie_1976}) it applies to many physical phenomena, producing what is widely regarded as the overwhelming precision of thermodynamics.  If valid, the ``$N^{ - 1/2}$ rule" applies only as $N \to \infty$; at very small $N$, a second effect must also be considered. 

For the canonical and other ensembles, the variability of the (superset) $\{ f_{ri} \}$ within {\it each} ensemble member is examined by (see above references):
\begin{equation}
\delta (f_r ) = \frac{{\sqrt {{\rm var} (f_r )} }}
{{\left\langle {f_r } \right\rangle }} = \frac{{\sqrt {\left[ {\left\langle {f_r ^2 } \right\rangle  - \left\langle {f_r } \right\rangle ^2 } \right]} }}
{{\left\langle {f_r } \right\rangle }}
\label{eq78}
\end{equation}
whence from (\ref{eq50})-(\ref{eq51}) and (\ref{eq71}):
\begin{equation}
\delta (f_r ) = \frac{1}
{{\left\langle {f_r } \right\rangle }}\sqrt { - \frac{{\partial \left\langle {f_r } \right\rangle }}
{{\partial \lambda _r }}}  = \frac{1}
{{\left\langle {f_r } \right\rangle }}\sqrt {\frac{{\partial ^2 \lambda _0 }}
{{\partial \lambda _r ^2 }}}  = \frac{1}
{{\left\langle {f_r } \right\rangle }}\sqrt { - \frac{{\partial ^2 \Phi ^*}}
{{\partial \lambda _r ^2 }}} 
\label{eq79}
\end{equation}
Whether or not this vanishes as $N \to \infty $ depends on the physical variable $r$ and the importance of interactions \cite[c.f. previous footnote]{Fowler_1936, Tolman_1938, Hill_1956, Davidson_1962}.  The variability of $\{ f_{ri} \}$ for the {\it total} ensemble can be examined using $\delta (\mathbb{N}f_r )$, where $\mathbb{N}$ is the number of ensemble members, giving a relation analogous to (\ref{eq77}).  It is commonly asserted that $\mathbb{N} \to \infty$ (e.g.\  \cite{Schrodinger_1952}), a rather questionable assumption. 
If correct, the total ensemble will be heavily concentrated at its ensemble means $\left\langle {f_r } \right\rangle, \forall r$.

Jaynes' \cite{Jaynes_1968, Jaynes_1978, Jaynes_1979} entropy concentration theorem considers the relative importance of the equilibrium probability distribution ${\mathbf p^*}=\{ p_i^* \}$ and some other distribution ${\mathbf p'}=\{ p_i' \}$.  From (\ref{eq34}) or (\ref{eq62}), the ratio of the probability of occurrence of ${\mathbf p^*}$ to that of ${\mathbf p'}$ is:
\begin{equation}
\frac{{\mathbb{P}^*}}{{\mathbb{P}'}} = \exp [N( - D^* + D')]
\label{eq80}
\end{equation}	
where ${\mathbb{P}^*}$, ${\mathbb{P}'}$ are the governing probability distributions and $D^*$, $D'$ are the cross-entropies corresponding  respectively to ${\mathbf p^*}$ and ${\mathbf p'}$.  This was originally formulated as the ratio of the number of ways in which ${\mathbf p^*}$ and ${\mathbf p'}$ can be realized \cite{Einstein_1905, Jaynes_1968}:
\begin{equation}
\frac{{\mathbb{W}^*}}{{\mathbb{W}'}} = \exp [N(H^* - H')]
\label{eq81}
\end{equation}	
where ${\mathbb{W}^*}$, ${\mathbb{W}'}$ are the weights and $H^*$, $H'$ are the entropies corresponding to ${\mathbf p^*}$ and ${\mathbf p'}$.  As shown by Jaynes \cite{Jaynes_1968, Jaynes_1978, Jaynes_1979}, for $N \to 1000$ even a small difference in $H$ gives an enormous ratio, revealing the combinatorial dominance of the maximum entropy position.

Assuming ${\mathbf p^*}$, ${\mathbf p'}$ satisfy the constraints ((\ref{eq28})-(\ref{eq29})), and taking the Stirling limits $N \to \infty$ and $n_i \to \infty$, an analysis similar to Kapur \& Kesavan \cite[\S2.4.6]{Kapur_K_1992} yields:
\begin{equation}
 - D^* + D' = H^* - H' = \sum\limits_{i = 1}^s {p_i '\ln \left( {\frac{{p_i '}}{{p_i ^*}}} \right)} 
\label{eq82}
\end{equation}
i.e. simply the directed divergence of ${\mathbf p'}$ from ${\mathbf p^*}$, from which ${\mathbf q}$ vanish (being incorporated into ${\mathbf p^*}$).  Eqs.\ (\ref{eq80})-(\ref{eq81}) then give:
\begin{equation}
\frac{{\mathbb{P}^*}}
{{\mathbb{P}'}} = \frac{{\mathbb{W}^*}}
{{\mathbb{W}'}} = \exp \left\{ {N\sum\limits_{i = 1}^s {p_i '\ln \left( {\frac{{p_i '}}
{{p_i ^*}}} \right)} } \right\}
\label{eq83}
\end{equation}
If we now put $p_i ' = p_i ^*(1 + \varepsilon _i )$, take a series expansion of $\ln p_i '$ about $\varepsilon _i  = 0$, and discard all polynomial terms higher than $\varepsilon _i ^2 $, it is shown by Kapur \& Kesavan \cite[\S2.4.7]{Kapur_K_1992} that (a quite different derivation is given by Jaynes \cite{Jaynes_1979}):
\begin{equation}
- D^* + D' = H^* - H' \approx \frac{1}
{2}\sum\limits_{i = 1}^s {\frac{{(p_i ' - p_i ^*)^2 }}
{{p_i ^*}}}  = \frac{1}
{{2N}}\sum\limits_{i = 1}^s {\frac{{(n_i ' - n_i ^*)^2 }}
{{n_i ^*}}}  = \frac{1}
{{2N}}\chi ^2 
\label{eq84}
\end{equation}
where $n_i ' = p_i 'N$ is the number of entities in state $i$ due to ${\mathbf p'}$; $n_i ^* = p_i ^*N$ is the expected number of entities in state $i$; and we recognize $\chi ^2$ as the chi-squared distribution of statistics \cite{Pearson_1900, Fisher_1922, Fisher_1924, Fisher_1925}.  In other words, we can determine the ``goodness of fit" of a distribution ${\mathbf p'}$ - or of some function $F(p)$ which generates ${\mathbf p'}$ - to a multinomial system, by comparing the calculated $\chi ^2$ to the table value $\chi ^2 (\nu ,1 - \alpha)$, where $\nu  = s - R - 1$ is the number of degrees of freedom and $\alpha$ is the significance level (upper tail or rejection area) \cite{Jaynes_1979}.  

As is well known \cite{Fisher_1924, Hoel_1962, Wise_1963, Kapur_S_1969} and dramatically illustrated by Jaynes \cite[chap 9]{Jaynes_2003}, the $\chi ^2$ statistic is an unreliable test for goodness of fit, being highly (and erroneously) sensitive to the occurrence of unlikely events.  There is no need to conduct the simplification of (\ref{eq84}); instead, from (\ref{eq82}):
\begin{equation}
 - D^* + D' = H^* - H' = \frac{1}{N}\sum\limits_{i = 1}^s {n_i '\ln \left( {\frac{{n_i '}}
{{n_i ^*}}} \right) = } \frac{\eta }{N}
\label{eq85}
\end{equation}
where $\eta$ is the correct test statistic for the goodness of fit of ${\mathbf p'}$ or its generator $F(p)$ to a multinomial system, subject to the Stirling limits ($\eta$ is given by Hoel \cite[\S10.1]{Hoel_1962}; and by Jaynes \cite[\S9.11.1]{Jaynes_2003} in the form $\psi  = 10\eta /\ln (10)$, using an obscure decibel notation.)  The calculated $\eta$ can be compared to the ``table value" $\eta (\nu ,1 - \alpha)$; alternatively, two distributions ${\mathbf p'}$ and ${\mathbf p''}$ can be ranked by comparing their corresponding $\eta'$ and $\eta''$.  Eqs.\ (\ref{eq83}) and (\ref{eq85}) finally give:
\begin{equation}
\frac{{\mathbb{P}^*}}{{\mathbb{P}'}} = \frac{{\mathbb{W}^*}}{{\mathbb{W}'}} = \exp (\eta ).
\label{eq86}
\end{equation}

\section{\label{sect3}APPLICABILITY OF MULTINOMIAL STATISTICS}
\subsection{\label{sect3_1}The ``Multinomial Family"}

Why have the Shannon information entropy and Kullback-Leibler cross-entropy proved to be of such utility, in an extremely wide range of disciplines?  The answer lies in the fact that an extraordinarily large number of probability functions $p_{i,...}$ or $p(x,...)$ of an observable, encompassing a wide range of statistical problems, can be obtained from the Stirling approximation to the multinomial distribution as special or limiting cases.  For example, in discrete statistics, the uniform, geometric, generalized geometric, power-function, Riemann zeta function, Poisson, binomial, negative binomial, generalized negative binomial and various Lagrangian distributions (and many others) have been obtained from the Shannon entropy subject to various constraints \cite{Kapur_1989b, Kapur_K_1992}.  Similarly, in continuous statistics, the uniform, normal (Gaussian), Laplace, generalized Cauchy, generalized logistic, generalized extreme value, exponential, Pareto, gamma, beta (of first or second kind), generalized Weibull, lognormal, Poisson, power-function and many new distributions, and various multivariate forms, can be obtained from the continuous form of the Shannon entropy subject to various constraints \cite{Kapur_1989b, Kapur_K_1992}.  Many additional distributions can be obtained from the Kullback-Leibler cross-entropy in discrete or continuous form, subject to various prior distributions and constraints \cite{Kapur_K_1992}.  All these functions therefore constitute particular examples of multinomial statistics, and collectively form the {\it multinomial family} of statistical distributions.  The broad applicability of the multinomial distribution, produced by the (fascinating) isomorphism of many probabilistic problems - such as of the ``balls-in-boxes" and ``multiple selection" systems described in \S \ref{sect2_3_1} - is responsible for the wide utility of the Kullback-Leibler cross-entropy and Shannon entropy functions.

\subsection{\label{sect3_2}Non-Multinomial Statistics}

Notwithstanding the success of multinomial statistics, it is important to emphasize that a number of statistical functions are incompatible with the Shannon entropy and/or Kullback-Leibler cross-entropy, and are therefore not of multinomial character.  Several of these (e.g.\  Bose-Einstein, Fermi-Dirac, R\'enyi, Tsallis and Kaniadakis entropies) reduce to the Shannon entropy as a limiting case \cite{Tolman_1938, Davidson_1962, Kapur_1989a, Renyi_1961, Tsallis_1988, Tsallis_2001, Kaniadakis_2001, Kaniadakis_2002}; such systems may therefore be approximated by multinomial statistics only when these limiting conditions are attained.  More thorough analyses of non-multinomial statistics must be deferred to later studies; however, their importance is here noted.

From the preceding analysis, it is clear that the definition of entropy (\ref{eq3}) promulgated by Boltzmann \cite{Boltzmann_1877} and Planck \cite{Planck_1901, Planck_1913} can be used irrespective of whether the distribution is of multinomial character.  A more comprehensive version, in which $\mathbb{P}$ now represents the governing probability distribution of {\it any} type and not only the multinomial distribution, is given in (\ref{eq39}).  The corresponding entropy is:
\begin{equation}
H({\mathbf{p}}) = K\left( {\frac{{\ln \mathbb{P}_{\mathbf{u}}}}{N} + C} \right) 
= K\left( {\frac{{\ln \mathbb{W}}}{N} + C'} \right)
\label{eq87}
\end{equation}
where $C$, $C'$ and $K$ are arbitrary constants.  (Note that the Boltzmann \cite{Boltzmann_1877} - Planck \cite{Planck_1901} formula (\ref{eq3}) is often misleadingly quoted as $S = k\ln \mathbb{W}$; this is correct only if $S$ refers to the {\it total} entropy of the system, not the entropy per unit entity.)  Indeed, it is not necessary to use a logarithmic transformation; for some distributions, some other transformation function $\phi$ may be more convenient, giving the {\it generalized definitions} of cross-entropy and entropy:
\begin{gather}
- D_{gen} ({\mathbf{p}},...|{\mathbf{q}},N,...) = \kappa ( {\phi (\mathbb{P},...) + C} ) 
\label{eq88} \\
H_{gen} ({\mathbf{p}},...|N,...) = \kappa ( {\phi (\mathbb{P}_{\mathbf{u}},...) + C}) = \kappa ( {\phi (\mathbb{W},...) + C'}) 
\label{eq89}
\end{gather}
with the only condition on $\phi$ being:
\begin{equation}
{{\rm extr}}\left[ {\phi (\mathbb{P},...)} \right] = {\rm max} \left[ {\mathbb{P},... } \right]
\label{eq90}
\end{equation}
where again $C$, $C'$ and $\kappa$ are arbitrary, whilst ``...''\ allows for other parameters or prior information. In many cases $\mathbb{P}$ will be a product-like function of $s$ local probability distributions $h_i(p_i,q_i,N,...)$; the appropriate choice of $\phi$ is the logarithm-like operator which transforms $\mathbb{P}$ neatly into a sum of terms in $\phi(h_i(p_i,q_i,N,...))$, simplifying its extremization\!\footnote{The recent derivation of the Tsallis \cite{Tsallis_1988} entropy by Suyari and co-workers \cite{Suyari_2004a, Suyari_2004b, Suyari_2005, Suyari_2006} using a transformation of the form $\phi=\ln_{2-q} (\mathbb{W}_{2-q})$, where $\ln_q$ is the $q$-logarithmic function and $\mathbb{W}_q$ is a $q$-multinomial coefficient, provides a fascinating example of an alternative transformation function.}  (for parallel discussions of deformed logarithms, see \cite{Naudts_2004, Kaniadakis_etal_2004}). Similarly, it may be convenient to choose $\phi$ and $\kappa$ which define an entropy function with a ``nice'' asymptotic limiting form, in the sense of large deviations theory \cite{Ellis_1985, Dembo_Z_1993}. Clearly, the information entropy (\ref{eq5}) given by Shannon \cite{Shannon_1948} - although derived from sound axiomatic postulates, and of quite broad scope - is strictly valid only for multinomial systems subject to the Stirling approximation.  This may be appropriate for communication signals of infinite length, but is surely insufficient to underpin the vast field of information theory in general.

\subsection{\label{sect3_3}Further Discussion}

In his many works, Jaynes expounds the ``Bayesian" or ``subjective" view of probabilities, which represent assignments of one's belief based on the available information, and argues against the ``frequentist" view in which probabilities are interpreted strictly as frequency assignments \cite{Jaynes_1957, Jaynes_1965, Jaynes_1968, Jaynes_1988}.  Separately, Jaynes demonstrates the equivalence of MaxEnt based on the Shannon entropy, and combinatorial analysis using the multinomial weight (the so-called Wallis derivation) \cite{Jaynes_1963, Jaynes_1968}.  At this point, however, he considers the combinatorial approach to represent a frequency interpretation, stating \cite{Jaynes_1968, Jaynes_2003}: ``the {\it probability} distribution which maximizes the entropy is numerically identical with the {\it frequency} distribution which can be realized in the greatest number of ways'' [his emphasis].  This identification of the combinatorial approach with the frequentist view is unfortunate; in fact, by applying MaxEnt based on the Shannon entropy, one {\it assumes} (implicitly) that the phenomenon being examined follows the multinomial distribution, and one uses one's prior knowledge to infer (hypothesize) the available states $i$ (for a parallel discussion, see Bhandari \cite{Bhandari_1976})\
\!\footnote{Jaynes appears to reach essentially this viewpoint in his final work \cite[chaps. 9, 11; especially \S 9.5-9.6, 11.4]{Jaynes_2003}.}. The calculated probability distribution $p_i^*$ is therefore valid only in the ``subjective" sense (i.e.\ exists only as an inference of the observer) until verified by experiment.  Even if so ``verified", there will always be room for doubt over its validity.

Indeed, the calculated MinXEnt probability (e.g.\  (\ref{eq25})-(\ref{eq26})) can be expressed in a reversed form of Bayes' theorem:
\begin{equation}
p_i^* = P^* (i|M,I) = \frac{{P(i|I)P^* (M|i,I)}}{{P^* (M|I)}} 
= \frac{{P(i|I)P^* (M|i,I)}}
{{\sum\limits_{i = 1}^s {P(i|I)P^* (M|i,I)} }}
\label{eq93}
\end{equation}
where $P$ is a probability, $P^*$ is a most probable (modal) probability, $i$ is the $i$th distinguishable outcome (datum) within a set of $s$ such outcomes, $M$ is the $i$th manifestation of the hypothesised model $\mathbb{P}$, not necessarily of multinomial form, and $I$ is the prior information. For the problems considered here, $I$ includes the constraints, any approximation or limit assumptions (e.g.\  the Stirling approximation) and any other relevant prior knowledge; if desired, these can be itemised separately. We immediately recognise the denominator in ({\ref{eq93}) as the partition function $Z_q$ (\ref{eq26}) or its equivalent, whilst $P(i|I)=P^*(i|I)=q_i$ is the prior probability. The generalized MinXEnt or MaxEnt methods therefore provide a method, in the absence of any sampling data, to ``bootstrap'' a sampling distribution $\{P^* (i|M,I)\}$ (we could call it the {\it posterior pre-sampling distribution}) from some hypothesis distribution $\{P^* (M|i,I)\}$ and prior distribution ${\mathbf q}$.  The latter two distributions are necessarily embedded within the governing distribution $\mathbb{P}$, being obtainable from it by extremization of (\ref{eq88}) or (\ref{eq89}) subject to $I$\
\!\footnote{A quite different connection between Bayesian and combinatorial perspectives was given recently \cite{Grendar_G_2004b}.}.

In consequence, the generalized definitions of cross-entropy and entropy given here ((\ref{eq88})-(\ref{eq89})) fit seamlessly into a Bayesian inferential framework \citep[c.f.][]{Cox_1961, Jaynes_1988, Tribus_1988}. In such cases, ${\mathbf q}$ represents a ``Bayesian prior distribution", ``Jeffrey's uninformative prior" \cite{Jeffreys_1932, Jeffreys_1961} or ``Jaynes' measure distribution" \cite{Jaynes_1963, Jaynes_1968}, whilst $\mathbb{P}$ represents one's postulated understanding of the probabilistic structure of the phenomenon at hand.  The broader Jaynesian program of maximum entropy analysis as a method of statistical inference is therefore untouched (in fact, enhanced) by the present analysis.  

Now considering the other bases of entropy listed in \S\ref{sect1}:
\begin{enumerate}
\renewcommand{\labelenumi}{(\alph{enumi})}
\item In the {\it inverse modelling basis} developed by Kapur, Kesevan and co-workers \cite{Kapur_K_1987, Kesavan_K_1989, Kapur_K_1992}, one works backwards from a hypothesized or observed probability distribution (${\mathbf p}$), prior distribution (${\mathbf q}$) and constraints (C0-C$R$), to obtain the measure of cross-entropy or entropy applicable to the process.  Using (\ref{eq88}) or (\ref{eq89}), such inverse methods could then be used to determine the governing probability distribution $\mathbb{P}$ of the process.  In effect, this approach was adopted in statistical mechanics by Planck \cite{Planck_1901}, Einstein \cite{Einstein_1905b, Einstein_1924, Einstein_1925} and Bose \cite{Bose_1924}, in which measurements of the thermodynamic entropy were used to determine the statistical weight. Alternatively, one can work ``sideways" from the observed, prior and governing distributions (${\mathbf p}$, ${\mathbf q}$ and $\mathbb{P}$) to determine the constraints \citep[c.f.][]{Kapur_etal_1995, Yuan_Kesavan_1998, Srikanth_2000}. Such methods offer powerful extensions to existing theory, yet have barely been examined in the literature. 
\item In the {\it game-theoretic basis} founded by von Neumann and Morgenstern \cite{vonNeumann_M_1944}, a decision tree or matrix is built up by analysis of a game between two or more players. The statistical structure of the game and optimal playing strategies are then determined. Originally devised for gambling and economics, this basis has found diverse application to political science, biological evolution, military strategy, counter-terrorism, queuing theory, operations research, system dynamics and many other fields. In many game-playing scenarios, there exists a game-theoretic equilibrium (in economics:\ ``Nash equilibrium'') at which no player can benefit by changing his/her strategy \cite{vonNeumann_M_1944, Nash_1950}; such equilibrium concepts have more recently been related to information measures, coding theory and MaxEnt \citep[e.g.][]{Topsoe_1993, Harremoes_T_2001, Topsoe_2002, Topsoe_2004, Grunwald_2004}. Although underpinned by axiomatic arguments, this basis is quite complementary to the combinatorial scheme outlined here, enabling the derivation of information measures and governing distributions for systems of complicated statistical or dynamic structure. 
\item In the {\it information-geometric} or {\it statistical manifold} basis, a statistical or probabilistic model is represented as a geometric structure (manifold) in some mathematical space, e.g.\ the space of probabilities, the space of population parameters or the space of possible distributions for the system \cite{Bhattacharyya_1943, Rao_1945, Amari_1985, Burbea_1986}. This structure can then be analysed geometrically, using measures of distance (metrics), shape and connectivity (topology), tangency and differentiability. As an example, a cross-entropy or divergence measure can be interpreted as the probabilistic distance of distribution ${\mathbf p}$ from ${\mathbf q}$ \cite{Bhattacharyya_1943, Kullback_L_1951, Kullback_1959, Burbea_1983}.  Information-geometric arguments have recently been applied to alternative entropy concepts \citep[e.g.][]{Vignat_P_2005}, but much more work is required in this field. Since the information-geometric basis serves as a representation of a system, rather than a cause, it is subsidiary to the other bases; however, it provides a valuable tool for the analysis of combinatorial concepts and distributions, and their connections to other physical theory (e.g.\ general relativity).  
\end{enumerate}

The analysis to this point has followed a long path, only to arrive more or less at its starting point:\ the combinatorial entropy of Boltzmann and Planck (although the idea is taken somewhat further than they had imagined).  The fact that this discussion is still necessary in the 21st century reflects the great gulf between present-day statistical mechanics and thermodynamics - still taught much as they were 50 or even 100 years ago - and the more recent but surprisingly narrow field of information theory initiated by Shannon \cite{Shannon_1948}.  The gulf persists despite the efforts of Bose, Einstein, Fermi and Dirac, amongst others, in statistical mechanics, and of Jaynes, Tribus, Kapur, Kesavan and many others in information theory and maximum entropy methods.  The two fields are, in fact, one.  Appreciation of this fact (by both sides) would permit the development of a much broader discipline of ``combinatorial information theory" than at present, applicable to many different types of problems. 

In our search for meaning in the universe and our existence, the sentiment of John Wheeler is frequently quoted:\ ``{\it it from bit}'' \cite{Wheeler_1990}; i.e.\ the entirety of existence ({\it it}) arises from information-theoretic principles ({\it bit}). This statement implies a universe built up by observer-participancy, individual measurement by measurement. However, from the present analysis \cite[see also][]{Niven_2005, Niven_2006}, each combinatorial family has a different kind of ``bit'', producing a vast array of different observational frameworks and observer-dependent (subjective?) realities. This can be paraphrased as ``{\it both it and bit from prob.}'', i.e.\ both existence and information theory arise from raw probabilistic constructs.  This principle clearly underpins the probabilistic definition of the second law of thermodynamics, ``{\it a system tends towards its most probable form}'', and may well explain the peculiarities of quantum mechanics \cite{Bose_1924, Einstein_1924, Einstein_1925, Fermi_1926, Dirac_1926, Tolman_1938, Davidson_1962, Niven_2005, Niven_2006}.

\section{\label{sect4}Conclusions}

The philosophical bases of the entropy and cross-entropy concepts are critically examined, with particular attention to the information-theoretic, axiomatic and combinatorial interpretations.  It is shown that the combinatorial basis, as first promulgated by Boltzmann and Planck, is the most fundamental (most primitive) of these three bases.  Not only does it provide (i) a derivation of the Kullback-Leibler cross-entropy and Shannon entropy functions, as simplified forms of the multinomial distribution subject to the Stirling approximation; the combinatorial approach also yields (ii) an explanation for the need to maximize entropy (or minimize cross-entropy) to find the most probable realization; and (iii) generalized definitions of entropy and cross-entropy  for systems which do not satisfy the multinomial distribution, i.e.\ which fall outside the domain of the Kullback-Leibler and Shannon measures. The information-theoretic and axiomatic bases of cross-entropy and entropy - whilst of tremendous importance and utility - are therefore seen as secondary viewpoints, which lack the breadth of the combinatorial approach.  The view of Shannon, Jaynes and their followers - in which the Shannon entropy or Kullback-Leibler cross-entropy is taken as the starting point and universal tool for analysis - is not seen as incorrect, but simply incomplete.  On the other hand, the viewpoint of many scientists - who consider statistical mechanics to be a branch of classical mechanics or quantum physics, rather than of statistical inference - is also incomplete.  A more detailed understanding of the combinatorial basis will enable development of a powerful body of ``combinatorial information theory", as a tool for statistical inference in all fields.  

The generic formulation of statistical mechanics developed by Jaynes  \cite{Jaynes_1957, Jaynes_1963, Jaynes_1968, Jaynes_2003, Tribus_1961a, Tribus_1961b, Kapur_1989b, Kapur_K_1987, Kapur_K_1992} is re-examined in light of the combinatorial approach. The analysis yields several new concepts including a generalized Clausius inequality, a generalized free energy (``free information") function, a generalized Gibbs-Duhem relation and phase rule, and a reappraisal of fluctuation theory and Jaynes' entropy concentration theorem. The free information concept provides a framework for the application of thermodynamic-like tools (e.g.\ state functions, cyclic integrals, efficiency ratios, Gibbs-Duhem and phase relations, Maxwell-like relations and Jaynes relations) for the analysis of probabilistic systems of {\it any} type. Finally, the combinatorial basis is shown to be embedded within a Bayesian statistical framework.


\section*{Acknowledgments}
This work began in 2002, and was in part completed during sabbatical leave in 2003 at Clarkson University, New York; McGill University, Quebec; Rice University, Texas and Colorado School of Mines, Colorado, supported by The University of New South Wales and the Australian-American Fulbright Foundation. The work benefited from valuable discussions with UNSW@ADFA colleagues, with participants at the 2005 NEXT Sigma Phi conference, Kolymbari, Crete, Greece, and (since 2006) with Marian Grendar.



\end{document}